\documentclass{article}

\usepackage{arxiv}

\usepackage[utf8]{inputenc} 
\usepackage[T1]{fontenc}    
\usepackage{hyperref}       
\usepackage{url}            
\usepackage{booktabs}       
\usepackage{amsfonts}       
\usepackage{nicefrac}       
\usepackage{microtype}      
\usepackage{lipsum}		
\usepackage{graphicx}
\usepackage{natbib}
\usepackage{doi}

\usepackage{amssymb}
\usepackage{latexsym}
\usepackage{amsmath}
\usepackage{subcaption}
\usepackage{float}
\usepackage{placeins}

\title{A Thermodynamically Consistent Manifold Model for Premixed Deflagrations \& Detonations}

\author{{\hspace{1mm}John B.~Boerchers}\thanks{Corresponding author, email: jb5027@princeton.edu} \\
	Department of Mechanical and Aerospace Engineering\\
	Princeton University\\
	Princeton, NJ \\
	\And
	{\hspace{1mm}Laura T. ~Thompson}\\
	Department of Mechanical and Aerospace Engineering\\
	Princeton University\\
	Princeton, NJ \\
	\And
	{\hspace{1mm}Matthew X. ~Yao}\\
	Department of Mechanical Engineering\\
	University of New Brunswick\\
	Fredericton, Canada \\
	Department of Mechanical and Aerospace Engineering\\
	Princeton University\\
	Princeton, NJ \\
	\And
	{\hspace{1mm}Michael E. ~Mueller}\\
	Department of Mechanical and Aerospace Engineering\\
	Princeton University\\
	Princeton, NJ \\
}

\renewcommand{\shorttitle}{A Thermodynamically Consistent Manifold Model}

\hypersetup{
pdftitle={A Thermodynamically Consistent Manifold Model for Premixed Deflagrations \& Detonations},
pdfsubject={q-bio.NC, q-bio.QM},
pdfauthor={John B.~Boerchers, Laura T.~Thompson, Matthew, X.~Yao, Michael E.~Mueller},
pdfkeywords={Manifold-based Modeling, Premixed Combustion, Detonation, Supersonic Combustion, Rotating Detonation Engine},
}

\begin{document}
\maketitle

\begin{abstract}
Accurate modeling of compressible premixed flames, encompassing both deflagrations and detonations, remains a significant challenge for predictive Large Eddy Simulation (LES) due to the strong coupling between the thermochemical state and the local thermodynamic state. This work presents a manifold-based turbulent combustion model that ensures a fully consistent thermodynamic state between model and flow solver through an iterative procedure. The framework reproduces critical quantities including temperature, radical species, and source term profiles, addressing limitations of existing approaches that rely on low-Mach perturbations or tabulated ZND detonations without thermodynamic consistency. Validation is performed against one-dimensional and high-fidelity RDE-like data, demonstrating that the thermodynamically consistent model consistently outperforms existing approaches across a broad range of compressible flame regimes --- including both deflagration and detonation. The results highlight the importance of fully accounting for the thermodynamic state to achieve accurate predictions. By capturing both deflagrative and detonative behavior within a single framework, the model provides a unified, versatile tool for LES of high-speed reacting flows and offers a foundation for future studies of compressible reacting flows, including applications to rotating detonation engines and other supersonic combustion systems.
\end{abstract}


\section{Introduction}
\label{sec1}
Recently, detonation-based propulsion and energy conversion devices have received renewed interest due to their theoretical thermodynamic advantages compared to their deflagration-based counterparts \cite{RamanRDE}. One such device is the rotating detonation engine (RDE), where one or several detonation waves travel around an annular passage, consuming a fuel–oxidzer mixture injected into the chamber and exhausting the combustion products through the exit plane. In particular, the pressure rise from the detonation allows for additional work to be extracted (compared to deflagration-based devices) resulting in an increased cycle efficiency.

However, despite the theoretical advantages, RDEs struggle to be practically realized due to the difficulty in controlling and maintaining stable detonation waves. Significant work has been devoted to investigating the coupling between the detonation characteristics (speed, stability, number of waves, etc.) and injector design \cite{Valencia2024}, wall heat flux \cite{Goto2022,Qiu_2024,Ishihara}, parasitic deflagration \cite{Chacon_Gamba_2019}, and exit plane characteristics \cite{Fotia2016}, among other considerations. While experiments continue to play an important role in furthering understanding of the fundamental processes, simulations have the advantage of providing access to the entire flow field and thermochemical state for analysis and design insights. However, given the limitations of current modeling approaches, simulations of three-dimensional RDEs with detailed chemistry \cite{Batista_Ross_Lietz_Hargus_2021,Prakash_Raman_Lietz_Hargus_Schumaker_2020} are extremely computationally expensive. In particular, Batista et al. \cite{Batista_Ross_Lietz_Hargus_2021} and Prakash \cite{Prakash_Raman_Lietz_Hargus_Schumaker_2020} et al.  considered gaseous methane-oxygen mixtures. The study from Batista \cite{Batista_Ross_Lietz_Hargus_2021} et al. mentions the use of six million CPU hours across 16,060 processors for a resolution of approximately five cells per induction length. Additionally, the resolution requirements for detonations are still subject to ongoing debate \cite{powers2005accurate}. The computational resources required for detailed RDE simulations greatly exceed those demanded by the iterative design procedure and motivates the development of accurate, low-cost models which can capture the necessary physics \cite{Gonzalez_Juez_Kerstein_Ranjan_Menon_2017}.

When analyzing the required model criteria, a particular requirement stands out: the need for a model capable of accurately predicting both deflagration and detonation. In the RDE, this criterion reveals itself in the crucial coupling between parasitic deflagration and detonation stability. In fact, the RDE is only one example of a process that includes both deflagrative and detonative combustion. Deflagrations are generally unavoidable in most detonation situations of practical interest. For example, deflagration-to-detonation transition (DDT) is another such process with important applications. Maintaining high fidelity is crucial to fully understanding the complex coupling between the large- and small-scale processes governing detonation wave stability and dynamics and reinforces the need for modeling improvement.

Manifold-based turbulent combustion models have demonstrated considerable success in reducing computational cost without significant loss in accuracy \cite{Peters_1984, Van_Oijen_2016}. These models rely upon the fact that the thermochemical state can be accurately represented as a function of a reduced set of scalars \cite{Pope_1992}.
As a result, the thermochemistry can be solved separately from the flow field (using the uniform background thermodynamic state and limited information about local mixing) and then accessed in the simulation as needed. Employing this approach greatly reduces computational cost by only requiring the solution of the transport equations for the reduced scalars instead of $k$ mass fractions $Y_k$ (where $k$ is an index denoting the chemical species which can vary from order $\mathcal{O}$ (10) - $\mathcal{O}$ (1000) depending on the chemical mechanism used \cite{Lu_Law_2009}).

However, these models usually consider only the low-Mach limit where the flow kinetic energy is negligible relative to the internal energy such that the thermodynamic state is unaffected by the flow dynamics.  In compressible flows, this decoupling no longer exists, and the background thermodynamic state can change locally due to the dynamics of the flow field (e.g., interconversion between kinetic energy and internal energy, shocks, etc.). 

Initial efforts to use manifold-based models for supersonic turbulent combustion did not directly incorporate compressibility into the model \cite{Oevermann}; rather, the approach (for nonpremixed flames in this case) used a baseline nonpremixed flame to compute species mass fractions $Y_k$ as a function of mixture fraction $Z$ and mixture fraction dissipation rate $\chi_{ZZ}$. In doing so, it is implicitly assumed that the fluctuating thermodynamic state has no effect on the composition. In other words, composition is computed at only a single pair of a nominal pressure $p$ and enthalpy $h$; in reality, these values exhibit significant local variation in supersonic flows. The coupling to the flow field is only done by computing the temperature from the transported energy using the composition from the model. Saghafian et al. \cite{Saghafian} takes this approach a step further to adapt it to a flamelet/progress variable (FPV) model. Similarly, species mass fractions are computed from a baseline solution and the results combined with the local transported energy to find temperature. However, since the progress variable source term is required for FPV, this approach applies a correction to the progress variable source term after computing the temperature to approximately account for the variation of the source term with density. Recently, Shunn et al. \cite{Shunn} utilized a similar approach, but modified for premixed flames, for use in detonations to study RDEs. To summarize, these approaches utilize a nominal background thermodynamic state to compute the low-Mach flame structure corresponding to their combustion mode of interest (nonpremixed or premixed). In doing so, there is no effect of compressibility on composition, and the flow field and the thermochemistry are only loosely coupled. These approaches can be classified as 'weakly-compressible' and they can be shown to exhibit errors of up to 500 K when applied to detonations \cite{Baumgart_Yao_Blanquart_2025}.

More recently, the compact memory representation of neural networks (NNs) has allowed more parameters to be incorporated into manifold models. For example, Demir et al. \cite{Demir_Kundu_Nunno_Som_Baurle_Drozda_2021} leverages a NN to additionally include pressure as a parameter in an FPV approach. However, this approach encountered coupling issues that will be addressed in this work. Specifically, the progress variable transport equation requires a more general description to incorporate changes in the equilibrium value resulting from the fluctuating background thermodynamic state. In fact, the approach from Demir et al. \cite{Demir_Kundu_Nunno_Som_Baurle_Drozda_2021} highlights an important point: adding manifold parameters alone is not sufficient; the model must also be consistently coupled to the flow solver.

For premixed detonations, Baumgart et al. \cite{Baumgart} argued that the baseline manifold solution must be representative of the physical problem and utilized one-dimensional detonation solutions as a function of progress variable and temperature constructed from a baseline Chapman-Jouguet (CJ) detonation with the added under- and over-driven counterparts. This approach changes the underlying flame structure and includes pressure variation across the solution, yet it still does not capture the correct background thermodynamic state and also cannot be applied to deflagrations.

Ultimately, all existing manifold-based approaches still fail to properly couple the thermodynamic state between the flow field and the model. This work hypothesizes that the primary difficulty inhibiting the success of compressible manifold-based combustion models is the need to match the thermodynamic states between model and flow solver. Recent work addressed this for nonpremixed flames by iterating upon the thermodynamic state from the flow solver and that provided by the model until convergence. This approach demonstrated that the consistent background thermodynamic state could be recovered within only a few iterations, yielding the first model capable of fully coupling the flow field and thermochemical state \cite{Cisneros_Garibay_Mueller_2024}. This work aims to extend the iterative approach to premixed deflagrations and detonations. 

The manuscript is organized as follows. Section \ref{premixed-combustion-model} will introduce the progress variable manifold-based turbulent combustion model within the Large Eddy Simulation (LES) framework, highlight the challenges associated with its application to compressible flows, and present the iterative approach for consistent thermodynamic coupling. Next, the performance of the model is evaluated, first against one-dimensional deflagrations and detonations in Sec. \ref{1d-analysis}, and subsequently against high-fidelity data from an RDE-like simulation in Sec. \ref{rde-analysis}, with results compared to other existing models.

\section{A Compressible Premixed Manifold-based Turbulent Combustion Model}
\label{premixed-combustion-model}
The model is presented in the context of LES, in which the governing equations are derived by applying a spatial filtering operation to the compressible Navier–Stokes equations for a reacting, Newtonian, ideal gas. Density-weighted (Favre) filtering is used for convenience in handling variable-density effects. Spatial filtering and Favre filtering are denoted by $(\bar{\cdot})$ and $(\tilde{\cdot})$ respectively. The filtered quantities are computed from the flow solver, and the subfilter combustion processes are modeled using the manifold model.

For adiabatic premixed combustion, the appropriate scalar for the manifold model is the progress variable $\Lambda \in [0,1]$, which typically uses a major product (or the sum of several) as a reference species to capture the progress of the reaction toward equilibrium:
\begin{equation}
\Lambda = \frac{Y_{R} - Y_{R,u}}{Y_{R,eq}-Y_{R,u}}.
\label{eq:lambda}
\end{equation}
$Y_R$ denotes the reference species, and $Y_{R,eq}$ and $Y_{R,u}$ denote the values of the reference species at equilibrium and the unburned state. This work will focus on hydrogen combustion and, as such, utilizes $Y_R$ = $Y_{\mathrm{H_2O}}$ as the reference species.

\subsection{Transport Equations}
\label{sec:transport}
The typical transported quantities for a compressible reacting flow are $\bar{\rho}$, $\bar{\rho}\tilde{u}_i$, $\bar{\rho}\widetilde{E}$, and $\bar{\rho} \widetilde{Y}_k$, where $\bar{\rho}$ is the density, $\tilde{u}_i$ the velocity in the $i$th direction, $\widetilde{E}$ ($=\tilde{e}+\frac{1}{2}\widetilde{u_k u_k}$) the total energy, and $\widetilde{Y}_k$ the mass fraction of species $k$. However, as mentioned previously, for a premixed manifold-based turbulent combustion model, the $k$ mass fraction transport equations will instead be replaced by the transport equations for progress variable $\widetilde{\Lambda}$ and its variance $\Lambda_v$, with the thermochemical state mapped to $\Lambda$. These equations are presented here in conservative form:

\begin{align}
\frac{\partial \bar{\rho}\widetilde{\Lambda}}{\partial t}
+ \frac{\partial \bar{\rho}\widetilde{\Lambda}\tilde{u}_j}{\partial x_j}
&= \frac{\partial}{\partial x_j}
  \left( \bar{\rho}\widetilde{D}
  \frac{\partial \widetilde{\Lambda}}{\partial x_j}\right)
  + \overline{\dot{m}}_{\Lambda_c}
  + \frac{\partial \tau_j^{sfs}}{\partial x_j},
\label{eq:prog-var-transport}
\\[10pt]
\frac{\partial (\bar{\rho}\Lambda_v)}{\partial t}
+ \frac{\partial (\bar{\rho} \Lambda_v \tilde{u}_j)}{\partial x_j}
&= \frac{\partial}{\partial x_j}
   \left(\bar{\rho} \widetilde{D}
   \frac{\partial \Lambda_v}{\partial x_j} \right)
   - \left[\bar{\rho}\widetilde{\chi}_{\Lambda \Lambda}
   - 2\bar{\rho}\widetilde{D}
     \frac{\partial \widetilde{\Lambda}}{\partial x_j}
     \frac{\partial \widetilde{\Lambda}}{\partial x_j} \right] \notag \\
&\quad - \frac{\partial}{\partial x_j} \Big[
   \bar{\rho}\widetilde{u_j\Lambda^2}
   - \bar{\rho}\tilde{u}_j\widetilde{\Lambda^2}
   - 2\bar{\rho}\widetilde{u_j\Lambda}\widetilde{\Lambda}
   + 2\bar{\rho}\tilde{u}_j\widetilde{\Lambda}^2 \Big] \notag \\
&\quad - 2\left(\bar{\rho}\widetilde{u_j \Lambda}
   - \bar{\rho}\tilde{u}_j \widetilde{\Lambda}\right)
   \frac{\partial \widetilde{\Lambda}}{\partial x_j}
   + 2 \Big( \overline{\Lambda \dot{m}_{\Lambda_c}}
   - \widetilde{\Lambda}\,\overline{\dot{m}}_{\Lambda_c} \Big),
\label{eq:prog-var-variance-transport}
\end{align}
where 
\begin{align}
\tau^{sfs}_j
&= -\bar{\rho}\left(\widetilde{\Lambda u_j} 
  - \widetilde{\Lambda} \widetilde{u}_j \right),
\label{eq:tau-sfs-u}
\\[4pt]
\widetilde{\chi}_{\Lambda \Lambda}
&= 2\widetilde{D\frac{\partial \Lambda}{\partial x_j}
  \frac{\partial \Lambda}{\partial x_j}},
\label{eq:sdr}
\end{align}
$D$ (=$\lambda/\rho c_p$) is the molecular diffusivity (defined for unity Lewis numbers), and $\dot{m}_{\Lambda_c}$ is the source term for a compressible progress variable.

The subfilter flux $\tau^{sfs}_{j}$, as well as the third and fourth terms on the right-hand-side of Eq. \ref{eq:prog-var-variance-transport}, can be closed using a dynamic Smagorinsky-like closure model \cite{Moin_dynamic}, and subfilter contributions to the molecular diffusion fluxes are ignored.  The second term on the right-hand-side of Eq. \ref{eq:prog-var-variance-transport} is the subfilter scalar dissipation rate and can be closed with a variety of models \cite{Raman_Pitsch_2007}.

One contribution of this work is the formulation of Eqs. \ref{eq:prog-var-transport} and \ref{eq:prog-var-variance-transport} in their generality for compressible flows --- in particular, their inclusion of compressibility effects in the progress variable source term $\dot{m}_{\Lambda_c}$. In a low-Mach flow, the constant background thermodynamic state  results in one unique value for the reference species at equilibrium $Y_{R,eq}$, and, with a fixed value for the unburned reference species $Y_{R,u}$, the only quantity in Eq. \ref{eq:lambda} that varies in space and time is $Y_R$. As a result, the species transport equation for $Y_R$ can be normalized by the constant $Y_{R,eq}-Y_{R,u}$ and filtered to obtain a transport equation for $\widetilde{\Lambda}$, which can be closed with commonly used models and then solved in LES.

However, the fluctuating thermodynamic state in a compressible flow field means that $Y_{R,eq}$ also varies in space and time. As a result, a more general formulation is needed for the $\widetilde{\Lambda} $ and $\Lambda_v$ transport equations. A derivation which allows $Y_{R,eq}$ to vary in space and time reveals the details in $\dot{m}_{\Lambda_c}$:

\begin{equation}
    \dot{m}_{\Lambda_c} = \frac{\rho \chi_{\Lambda Y_{R,eq}} + \dot{m}_{R} - \Lambda \dot{m}_{Y_{R,eq}}}{Y_{R,eq} - Y_{R,u}} 
    \label{eq:high-speed-source-term}
\end{equation}
where
\begin{equation}
    \chi_{\Lambda Y_{R,eq}} = 2D\frac{\partial \Lambda}{\partial x_j}\frac{\partial Y_{R,eq}}{\partial x_j},
    \label{eq:chi-ly}
\end{equation}
\begin{equation}
    \dot{m}_{Y_{R,eq}} = \frac{\partial \rho Y_{R,eq}}{\partial t} + \frac{\partial \rho u_j Y_{R,eq}}{\partial x_j} - \frac{\partial}{\partial x_j}\left(\rho D \frac{\partial Y_{R,eq}}{\partial x_j}\right),
    \label{eq:source-term-y-ref-eq}
\end{equation}
and $\dot{m}_R$ is the chemical source term for the reference species. While the low-Mach progress variable source term, $\dot{m}_{\Lambda} = \dot{m}_R/(Y_{R,eq}-Y_{R,u})$, remains unchanged, a source term now appears to capture the local changes in the value of the reference species at equilibrium $\dot{m}_{Y_{R,eq}}$. Additionally, an extra coupling term appears in the form of a new scalar dissipation rate $\chi_{\Lambda Y_{R,eq}}$. The compressible progress variable source term will be obtained using the manifold model and, as such, its closure will be presented following the discussion of premixed manifold-based combustion models in Sec. \ref{Compressible-prog-var}.

\subsection{Premixed Manifold-based Combustion Modeling}
The development in this work follows the notation and framework presented by Mueller \cite{Mueller2020}. In the context of premixed combustion with unity effective Lewis numbers, this framework provides equations for the thermochemical state ($h, Y_k$) as a function of $\Lambda$ and parameterized by $\chi_{\Lambda \Lambda}$:
\begin{align}
\frac{d Y_k}{d \Lambda}\dot{m}_{\Lambda} 
&= \frac{\rho \chi_{\Lambda \Lambda}}{2}\frac{d^2 Y_k}{d \Lambda^2} + \dot{m}_k,
\label{eq:manifold-yk}
\\[6pt]
\frac{d h}{d \Lambda} \dot{m}_{\Lambda} 
&= \frac{\rho \chi_{\Lambda \Lambda}}{2}\frac{d^2 h}{d \Lambda^2}.
\label{eq:manifold-h}
\end{align}
The progress variable dissipation rate $\chi_{\Lambda \Lambda}$ describes the degree of mixing and its influence on the thermochemistry. Its dependence on the progress variable $\chi_{\Lambda \Lambda}\left(\Lambda\right)$ must be modeled, and the reader is referred to Refs. \cite{Mueller2020,Nguyen_2010,Lacey_VanderKam_Sundaresan_Mueller_2024} for details on commonly adopted approaches.

These equations require the specification of two sets of boundary conditions: the species and enthalpy of the unburned state $Y_k(\Lambda=0)$ and $h(\Lambda=0)$ and the equilibrium state $Y_k(\Lambda=1)$ and $h(\Lambda=1)$. Provided the unburned state, the needed quantities at $\Lambda=1$ are determined from chemical equilibrium at fixed $h, p$. The corresponding temperatures can be backed out from $Y_k$ and $h$. Using these boundary conditions, the solution to these equations is an isobaric premixed flame; in other words, it is a fixed $h, p$ flame.

Once $Y_k(\Lambda)$ and $h(\Lambda)$ are known through the solution of the manifold equations, their solution can be convolved against a presumed subfilter PDF $\widetilde{P}$ to provide the filtered value for an arbitrary thermochemical quantity $\widetilde{\phi}$ needed in LES:
\begin{equation}
    \widetilde{\phi}=\int\phi(\Lambda;\chi_{\Lambda \Lambda})\widetilde{P}(\Lambda;\widetilde{\Lambda}, \Lambda_v)d\Lambda
    \label{eq:pdf-convolution}.
\end{equation}
The PDF is typically presumed to be a beta distribution  parametrized by $\widetilde{\Lambda}$ and $\Lambda_v$ \cite{Bray_Champion_Libby_Swaminathan_2006}.

In low-Mach reacting flows the approximately uniform background thermodynamic state and the associated boundary conditions ($Y_k$ and $h$ at $\Lambda =0$ and $\Lambda = 1$) and thermodynamic pressure required for Eqs. \ref{eq:manifold-yk} and \ref{eq:manifold-h} do not change. This decoupling is, in fact, the motivation for manifold-based combustion models, which can separately compute the thermochemical state as a function of $\Lambda$ and the coupling to the flow field is reasonably approximated using only $\chi_{\Lambda \Lambda}$.

Conversely, for compressible flows the interplay between the flow field dynamics, the thermodynamic state, and the chemistry is entirely coupled. Quantitatively, this manifests in the boundary conditions for $h$ at $\Lambda=0$ and the pressure. These are required to compute the thermochemical state and should come from the flow solver; however, the flow solver requires knowledge of the thermochemical state to evaluate the equation of state, creating a circular dependence. Consequently, manifold-based combustion models have seen limited applicability in compressible and supersonic reacting flows.

To address this, this paper adapts the nonpremixed approach from Cisneros-Garibay and Mueller \cite{Cisneros_Garibay_Mueller_2024} for compressible and supersonic premixed flames to fully incorporate the thermodynamic state. The presented approach solves the manifold equations as they are presented in Eqs. \ref{eq:manifold-yk} and \ref{eq:manifold-h}, maintaining the structure of the fixed $(h,p)$ solution, but incorporating the fluctuating background thermodynamic state from the flow solver by iterating upon $h_{\Lambda=0}$ and $p$. It will be shown in Sec. \ref{flame-structure} that correctly capturing the background thermodynamic state is critical to obtaining an accurate thermochemical state.

\begin{figure*}[!t]
\centering
\includegraphics[scale=0.38]{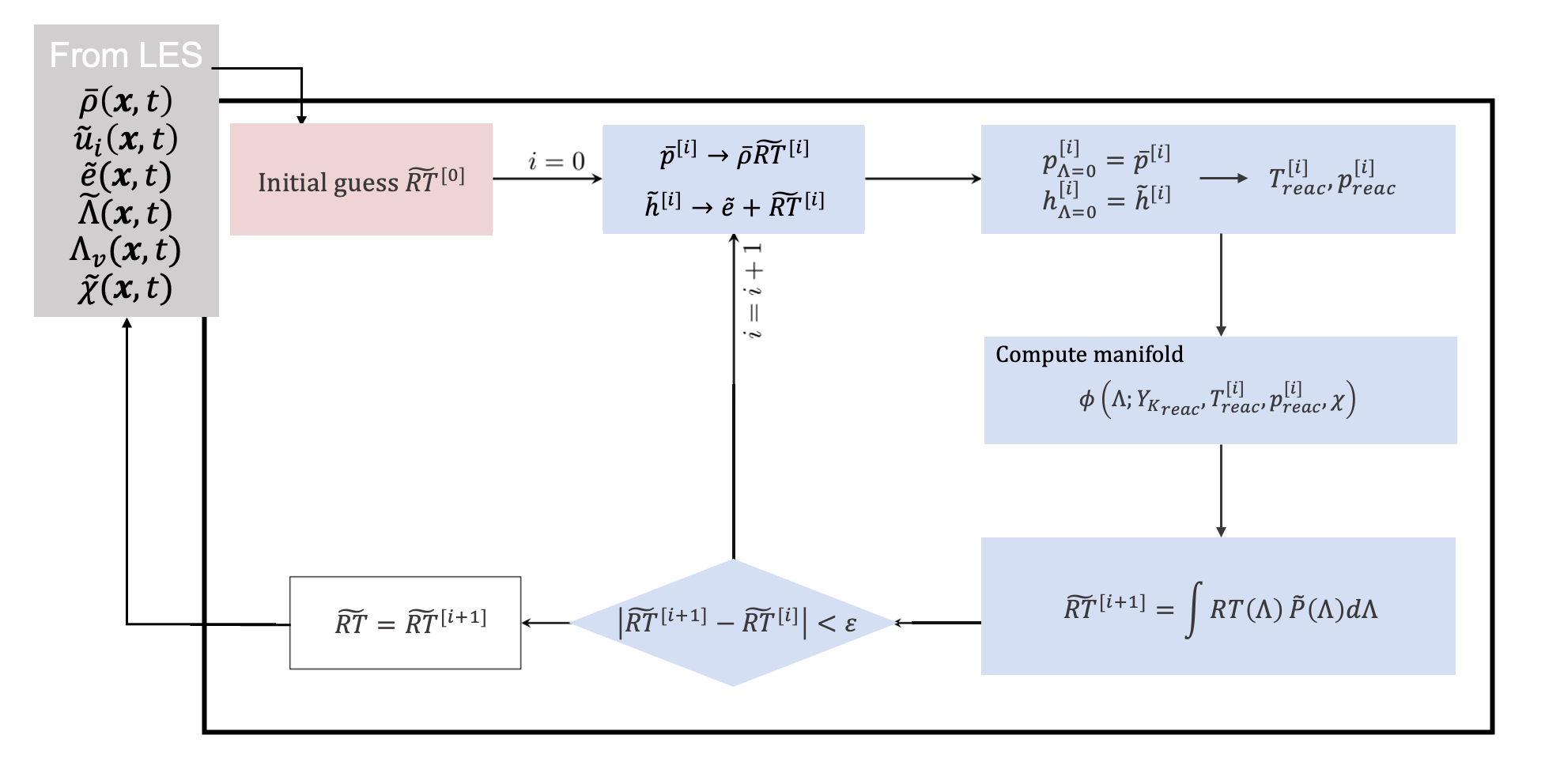}
\caption{Iterative algorithm for incorporating the fluctuating background thermodynamic state into the manifold model. Figure adapted from Ref. \cite{Cisneros_Garibay_Mueller_2024} but recast for premixed combustion. \label{fig:algorithm}}
\end{figure*}

Figure \ref{fig:algorithm} demonstrates this in more detail. First, an initial guess is provided for $\widetilde{RT}$. This is obtained by solving Eqs. \ref{eq:manifold-yk} and \ref{eq:manifold-h} with a guess for the background state (say, for example, $h_{\Lambda=0}$ corresponding to $T_{u}=300 K$ and $p=1$ atm) and obtaining the corresponding $\widetilde{RT}$ from the solution. Once this guess is provided, $\bar{p}$ and $\tilde{h}$ can be computed for an ideal gas using the state equations $\bar{p}=\bar{\rho}\widetilde{RT}$ and $\tilde{h}=\tilde{e}+\widetilde{RT}$ in combination with the transported $\bar{\rho}$ and $\widetilde{e}$. Next, the unburned boundary condition can be obtained utilizing the fact that the underlying flame structure is treated as isobaric with fixed ($h$, $p$). As a result, $\tilde{h}=h(\Lambda=0)=h(\Lambda=1)$. Similarly, the pressure is treated as a constant and $\bar{p} = p(\Lambda)$. The equilibrium boundary $\Lambda=1$ can be determined by computing chemical equilibrium for this $h,p$. Now, Eqs. \ref{eq:manifold-yk} and \ref{eq:manifold-h} can be solved and the resulting thermochemical state determined. The resulting $RT(\Lambda)$ profile is convolved against the presumed subfilter PDF $\widetilde{P}(\Lambda)$ to obtain a new $\widetilde{RT}$. The difference between the new $\widetilde{RT}^{[i+1]}$ and the previous $\widetilde{RT}^{[i]}$ is compared, and, if the result is less than a user defined tolerance, $\epsilon$, then the algorithm has converged; otherwise, the new $\widetilde{RT}$ is used for the next iteration and the cycle continues.

Equations \ref{eq:manifold-yk} and \ref{eq:manifold-h} can be solved beforehand and tabulated as a function of $\Lambda$ with parameters $\chi_{\Lambda \Lambda}$, $p$, and $h$ for use in LES, from which the full thermochemical state (mass fractions $Y_K$, enthalpy $h$, temperature $T$) and any temperature-dependent properties such as transport coefficients can be obtained. However, precomputing the entire range of solutions over the multi-dimensional space can introduce large memory requirements or interpolation errors. These issues can be greatly alleviated using an alternative approach known as In-Situ Adaptive Manifolds (ISAM) \cite{Lacey_Novoselov_Mueller_2021}  which solves the equations in-situ and stores and accesses solutions only when needed, avoiding the need to precompute all solutions and minimizing memory requirements by accessing only the set of solutions needed for the simulation.

\subsection{Closure for the Compressible Progress Variable Source Term}
\label{Compressible-prog-var}
Following the determination of the thermochemical state, the progress variable source term $\dot{m}_{\Lambda_c}$ can be obtained from the manifold model and convolved against $\tilde{P}$ to obtain the filtered value needed in Eqs. \ref{eq:prog-var-transport} and \ref{eq:prog-var-variance-transport} for the next time step in LES.
Since $\dot{m}_{\Lambda_c}$ is obtained using the manifold model, the terms within that depend on time and space must be addressed. The temporal and spatial derivatives for $Y_{R,eq}$ can be computed by applying the state principle, expanding them as functions of two state variables and the chemical species mass fractions. Assuming fixed $h, p$ flames simplifies the closure. In this approach, the progress variable requires the $Y_{R,eq}$ obtained \emph{isobarically} (fixed $h$ and $p$) from the local background thermodynamic state (say local $e$, $\rho$, and $Y_k$), uniquely defining the trajectory from unburned to equilibrium and allowing parameterization by $e$, $\rho$, and $Y_{k,u}$. Since $Y_{k,u}$ are constant, its derivatives vanish, eliminating the final terms in Eqs. \ref{eq:expansion1-notsimplified} and \ref{eq:expansion2-notsimplified}:

\begin{equation}
    \frac{\partial Y_{R,eq}}{\partial t} =
        \frac{\partial e}{\partial t} \frac{\partial Y_{R,eq}}{\partial e}
        + \frac{\partial \rho}{\partial t} \frac{\partial Y_{R,eq}}{\partial \rho}
        + \sum_{k=1}^{N_{spec}} \frac{\partial Y_{k,u}}{\partial t} \frac{\partial Y_{R,eq}}{\partial Y_{k,u}},
    \label{eq:expansion1-notsimplified}
\end{equation}

\begin{equation}
    \frac{\partial Y_{R,eq}}{\partial x_j} =
        \frac{\partial e}{\partial x_j} \frac{\partial Y_{R,eq}}{\partial e}
        + \frac{\partial \rho}{\partial x_j} \frac{\partial Y_{R,eq}}{\partial \rho}
        + \sum_{k=1}^{N_{spec}} \frac{\partial Y_{k,u}}{\partial x_j} \frac{\partial Y_{R,eq}}{\partial Y_{k,u}}.
    \label{eq:expansion2-notsimplified}
\end{equation}
$\partial Y_{R,eq}/\partial e$ and $\partial Y_{R,eq}/\partial \rho$ are obtained numerically by perturbing $e$ and $\rho$ individually using a chemical equilibrium solver such as CEQ \cite{CEQ_Pope_2004}. The terms comprising Eq. \ref{eq:source-term-y-ref-eq} are explored further in the analysis presented in Appendix~\ref{sec:source-term}. This analysis finds that the coupling scalar dissipation rate $\chi_{\Lambda Y_{R,eq}}$ is very small relative to the other terms in Eq. \ref{eq:high-speed-source-term} and the diffusion term is very small relative to the temporal and convective terms in Eq. \ref{eq:source-term-y-ref-eq}. Therefore, these terms can be neglected. In summary, with the simplifications above, $\dot{m}_{\Lambda_c}$ is then:

\begin{equation}
\dot{m}_{\Lambda_c}
=
\frac{1}{Y_{R,\mathrm{eq}} - Y_{R,u}}
\left[
    \dot{m}_R
    - \Lambda \rho
    \left(
        \frac{\partial Y_{R,\mathrm{eq}}}{\partial e}
        \frac{De}{Dt}
        +
        \frac{\partial Y_{R,\mathrm{eq}}}{\partial \rho}
        \frac{D\rho}{Dt}
    \right)
\right].
\label{eq:final-compressible-prog}
\end{equation}
With this, Eqs. \ref{eq:prog-var-transport} and \ref{eq:prog-var-variance-transport}, using the above source term from Eq. \ref{eq:final-compressible-prog}, constitute a closed set of equations for the compressible $\widetilde{\Lambda}$ and $\Lambda_v$ needed for the algorithm.

\section{One-dimensional Analysis}
\label{1d-analysis}
To evaluate the model, two representative one-dimensional premixed flames are considered: a compressed deflagration and a Zeldovich–Neumann–Döring (ZND) detonation. Additional cases (a weakly-compressed deflagration and an expanded deflagration) are available in Appendix~\ref{sec:additional-one-dimensional-analysis}. All cases consider a stoichiometric H$_2$-air mixture ($\phi=1$) for the unburned $Y_k$ and employ the H$_2$-air mechanism from Li et al. \cite{Li_Zhao_Kazakov_Dryer_2004}. Additionally, the Shock and Detonation Toolbox was used for obtaining the ZND profile \cite{edl_sdtoolbox_2023}. The deflagration was computed using the code PDRs detailed in Ref. \cite{Mueller2020}, which solves Eqs. \ref{eq:manifold-yk} and \ref{eq:manifold-h}. To obtain the unburned state for the compressed case, the stoichiometric H$_2$-air mixture was isentropically compressed from $T=300$ K and $p=1$ atm. The ZND case was initialized with the same stoichiometric H$_2$-air mixture at $T=300$ K and $p=1$ atm. The upstream velocity was set to the Chapman-Jouguet (CJ) velocity, yielding a post-shock state (Neumann state) $T\approx1540$ K and $p\approx27.9$ atm. All plots in this section show the ZND structure starting from the post-shock state. The resulting unburned states are displayed in Table \ref{table:unburned}. Both cases in this analysis use the $\chi_{\Lambda \Lambda}(\Lambda)$ model provided by Nguyen et al. \cite{Nguyen_2010} with a reference progress variable dissipation rate $\chi_{\Lambda \Lambda,ref}$ of 25000 s$^{-1}$. Henceforth, analysis will examine only the chemical component of the source term to facilitate comparison to other models and focus on the resulting thermochemical state. 

\begin{table}[!t]
    \caption{\label{table:unburned}Unburned state for the one-dimensional cases: (a) compressed deflagration and a (b) ZND detonation. Each case uses a stoichiometric H$_2$–air mixture ($\phi=1$).}
    \centering
    \renewcommand{\arraystretch}{1.3}
    \fontsize{8pt}{9pt}\selectfont
    \begin{tabular*}{\columnwidth}{@{\extracolsep{\fill}} l c c c c c }
    \hline
    \textbf{Case} & T (K) & p (kPa) & \textbf{$Y_{\rm H_2}$} & \textbf{$Y_{\rm O_2}$} & \textbf{$Y_{\rm N_2}$} \\
    \hline
    (a) & 515 & 668   & 0.029 & 0.226 & 0.745 \\
    (b) & 1540 & 2824 & 0.029 & 0.226 & 0.745 \\
    \hline
    \end{tabular*}
\end{table}

\subsection{Perturbation Model Overview}
The results of the one-dimensional analysis will be compared to the Shunn et al. \cite{Shunn} model; this model is an extension of the Sagahafian et al. \cite{Saghafian} model for detonations. In this model, one baseline solution is computed using a nominal background thermodynamic state, and then fitting parameters are calibrated in advance using a second perturbed solution at a different thermodynamic state. First, the ratio of specific heats $\gamma$ is fit as a linear function of temeperature between the baseline and perturbed solutions:
\begin{equation}
    \gamma(T) = \gamma_0 + a_{\gamma}(T - T_0).
    \label{eq:fitting-1}
\end{equation}
In Eq. \ref{eq:fitting-1}, $T_0$ and $\gamma_0$ are the temperature and specific heat ratio of the baseline solution. The fitting parameter $a_{\gamma}$ is selected such that $\gamma(T)$ matches the specific heat ratio of the perturbed solution at a given (unnormalized) progress variable $C$.

Next, the progress variable source term $\dot{m}_C(\rho, T)$ is fit from the baseline progress variable source term $\dot{m}_{C,0}$ at the baseline density $\rho_0$ and temperature $T_0$ using an Arrhenius-like expression to obtain the fitting parameters $a_{\rho}$ and $T_a$:
\begin{equation}
    \frac{\dot{m}_C}{\dot{m}_{C,0}} = \left(\frac{\rho}{\rho_0}\right)^{a_{\rho}} exp\left[-T_a\left(\frac{1}{T}-\frac{1}{T_0}\right)\right].
    \label{eq:fitting-3}
\end{equation}
With $a_{\gamma}$, $a_{\rho}$, $T_a$ calibrated, the model can now be applied to estimate temperature. Using the linear fit for the specific heat ratio and assuming the composition does not change due to compressibility effects, the temperature can be computed using the following relationship:
\begin{equation}
    T = T_0 + \frac{\gamma_{0}-1}{a_{\gamma}} \left[exp\left(\frac{a_{\gamma}(e-e_0)}{R_0}\right)-1\right].
    \label{eq:fitting-2}
\end{equation}
Then, the progress variable source term  $\dot{m}_C$ is estimated with the transported density $\rho$ and estimated temperature $T$ using Eq. \ref{eq:fitting-3} with the tuned parameters. Since this work focuses on hydrogen combustion and uses H$_2$O for the progress variable, $\dot{m}_C$ will henceforth be denoted, equivalently, as $\dot{m}_{{\rm H}_2{\rm O}}$.

\subsection{Compressed Deflagration}
\label{Compressed Deflagration}
\begin{figure*}[!t]
    \centering
    \begin{subfigure}{0.49\textwidth}
        \centering
        \includegraphics[width=\textwidth]{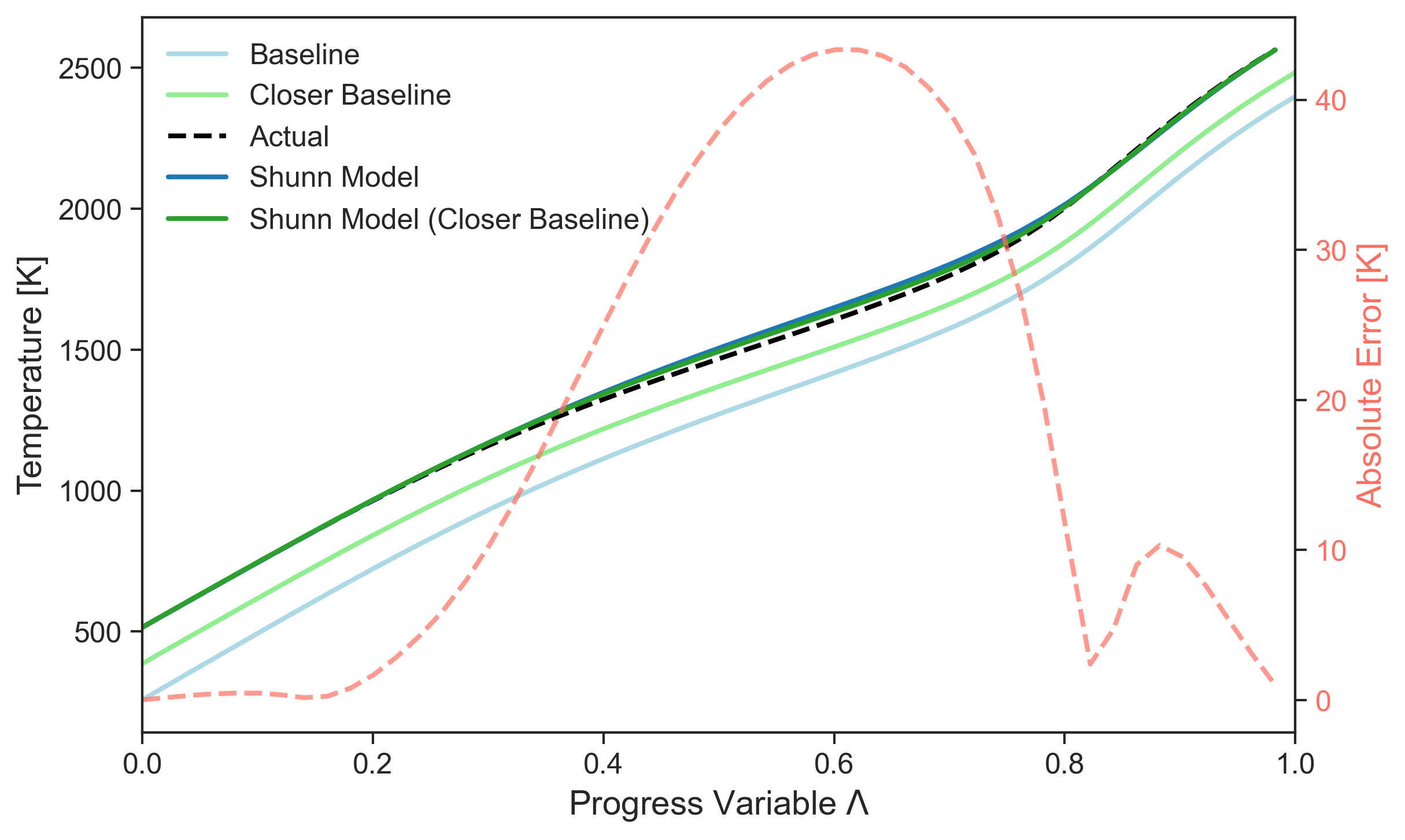}
        \caption{Temperature comparison}
        \label{fig:shunn-temp}
    \end{subfigure}
    \hfill
    \begin{subfigure}{0.49\textwidth}
        \centering
        \includegraphics[width=\textwidth]{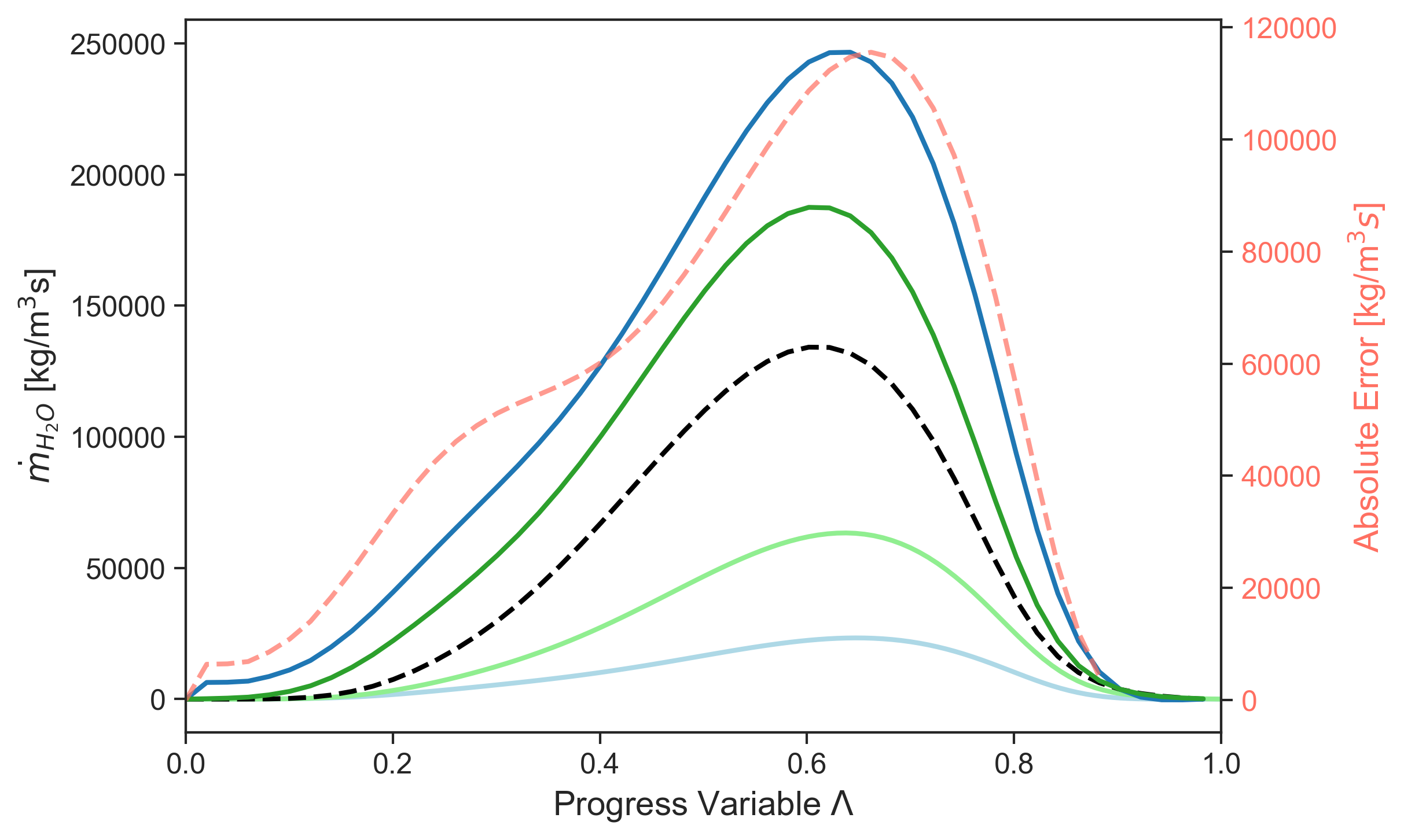}
        \caption{Source term comparison}
        \label{fig:shunn-src}
    \end{subfigure}

    \caption{Temperature and $\dot{m}_{{\rm H}_2{\rm O}}$ comparison for the Shunn et al. \cite{Shunn} model applied to the compressed deflagration case. The fitting parameters for the model are tuned at $\Lambda=0.5$, and two different baseline solutions are used to illustrate the effect of the initial profile on the results. The absolute error displayed corresponds to the baseline case.}
    \label{fig:Shunn-tuning}
\end{figure*}

\begin{figure*}[!t]
    \centering
    \begin{subfigure}[b]{0.49\textwidth}
        \centering
        \includegraphics[width=\textwidth]{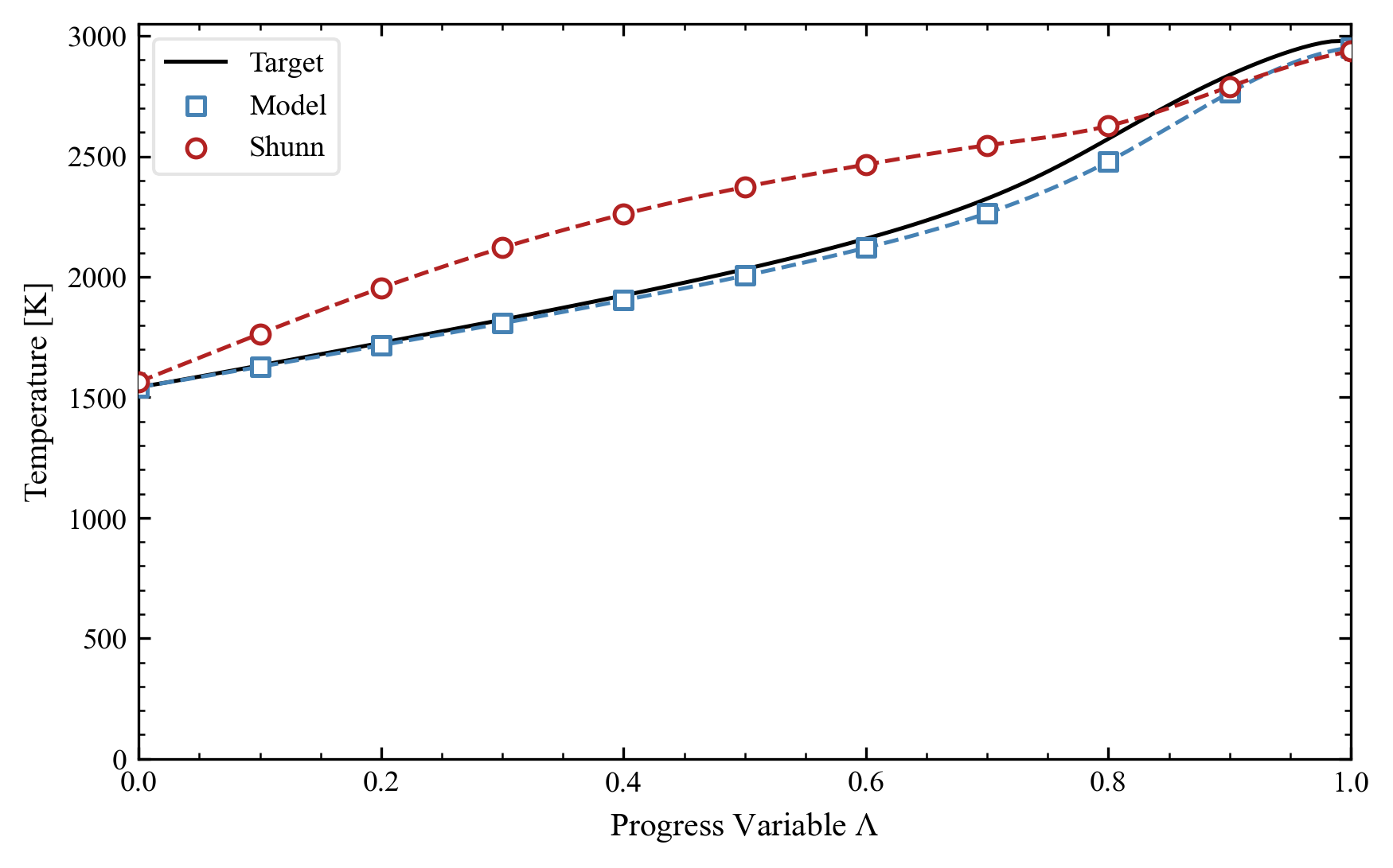}
        \caption{Temperature profile}
        \label{fig:znd-temp}
    \end{subfigure}
    \hfill
    \begin{subfigure}[b]{0.49\textwidth}
        \centering
        \includegraphics[width=\textwidth]{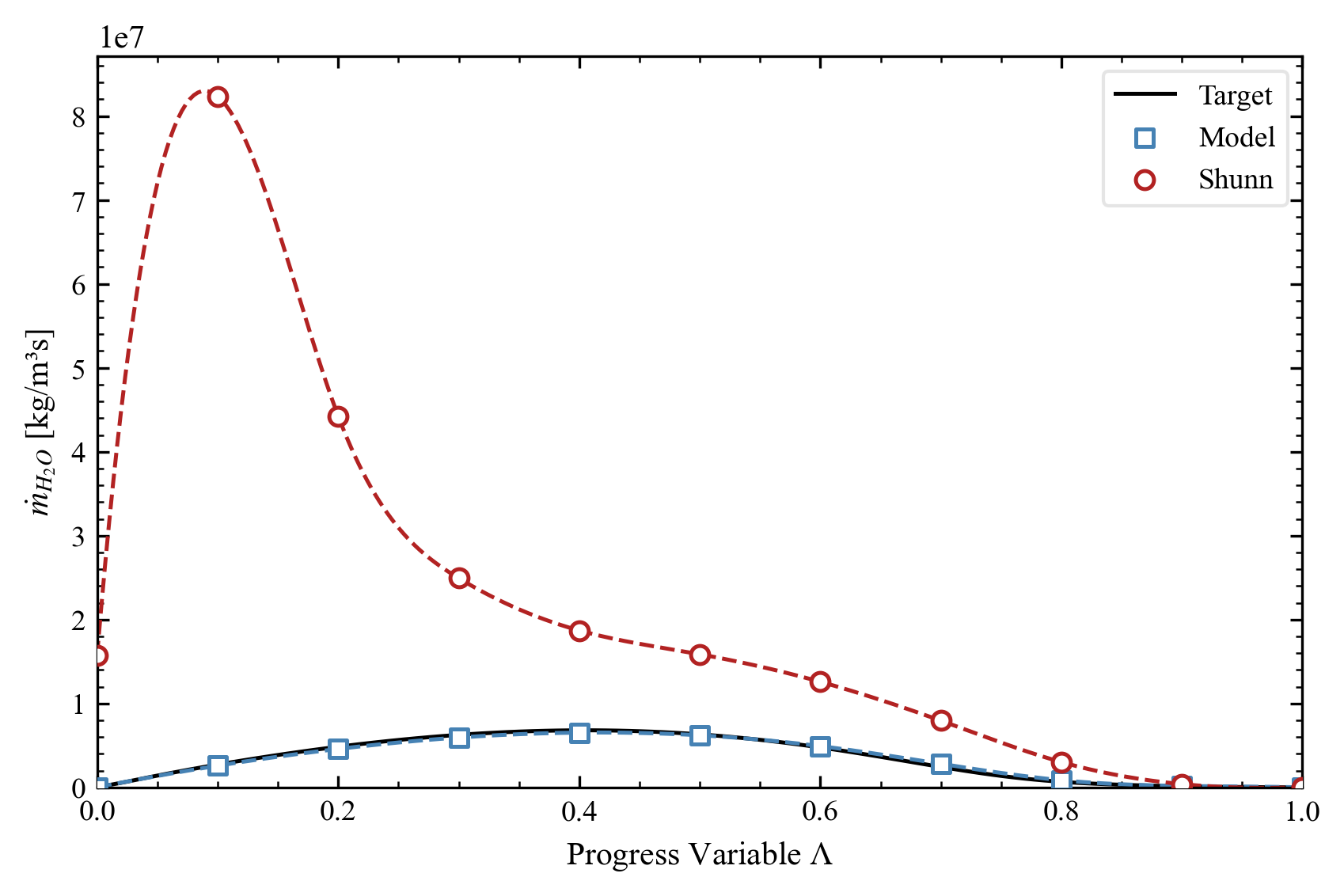}
        \caption{H$_2$O source term profile}
        \label{fig:znd-src}
    \end{subfigure}

    \caption{Convergence of the model for the ZND case at points from $\Lambda =$ 0.0 to 1.0 increasing in increments of 0.1. The thermodynamically consistent model is represented by the blue squares and the Shunn et al. \cite{Shunn} model by the red circles. Left: temperature; right: H$_2$O source term.}
    \label{fig:znd-case}
\end{figure*}

A stoichiometric H$_2$-air mixture with unburned $T=300$ K and $p=1$ atm was used for the baseline profile. The compressed deflagration (with the unburned state described in Table \ref{table:unburned}) was used for the perturbed profile. The Shunn \cite{Shunn} model was tuned at the same progress variable that is the target of the analysis ($\Lambda=0.5$). Figure \ref{fig:Shunn-tuning} shows the temperature (Fig. \ref{fig:shunn-temp}) and source term (Fig. \ref{fig:shunn-src}) profiles for a baseline from an unburned state with $T=300$ K and $p=101$ kPa (the nominal baseline), as well as a closer baseline with an unburned state of $T=386$ K and $p=501$ kPa to highlight the role of the baseline profile.
Figure \ref{fig:Shunn-tuning} also shows the predicted profiles for $T(\Lambda)$ and $\dot{m}_{{\rm H}_2{\rm O}}(\Lambda)$ for each corresponding baseline solution. At $\Lambda=0.5$, the Shunn model using the baseline deflagration overpredicts the temperature by approximately $37$ K, while the closer baseline reduces this overprediction to $25$ K. The source term demonstrates $\mathcal{O}(1)$ errors, with the baseline solution exhibiting a $\approx76$\% relative error, and, similarly, the closer baseline solution proving only slightly better with a relative source term error of approximately $42$\%. Accurate prediction of the reference species source term $\dot{m}_{{\rm H}_2{\rm O}}$ is a key model criterion, for errors in $\dot{m}_{{\rm H}_2{\rm O}}$ will propagate, producing increasingly incorrect $\Lambda$ values in subsequent timesteps. Additionally, since this model is founded upon a perturbation around a baseline solution, it is expected to perform increasingly worse the further the actual thermodynamic state is from the baseline state. In particular, the accuracy will worsen significantly for detonations, which will be explored in more detail shortly (see Fig. \ref{fig:znd-case}).

\begin{figure*}[!t]
    \centering
    \begin{subfigure}[b]{0.49\textwidth}
        \centering
        \includegraphics[width=\textwidth]{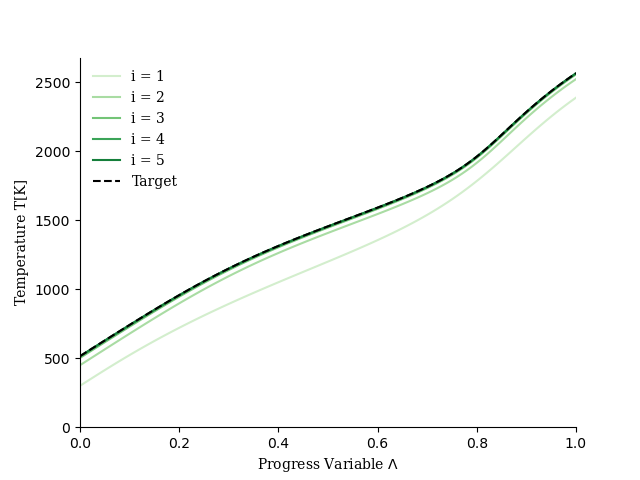}
        \caption{Temperature iterations}
        \label{fig:strongly-temp-iterations}
    \end{subfigure}
    \hfill
    \begin{subfigure}[b]{0.49\textwidth}
        \centering
        \includegraphics[width=\textwidth]{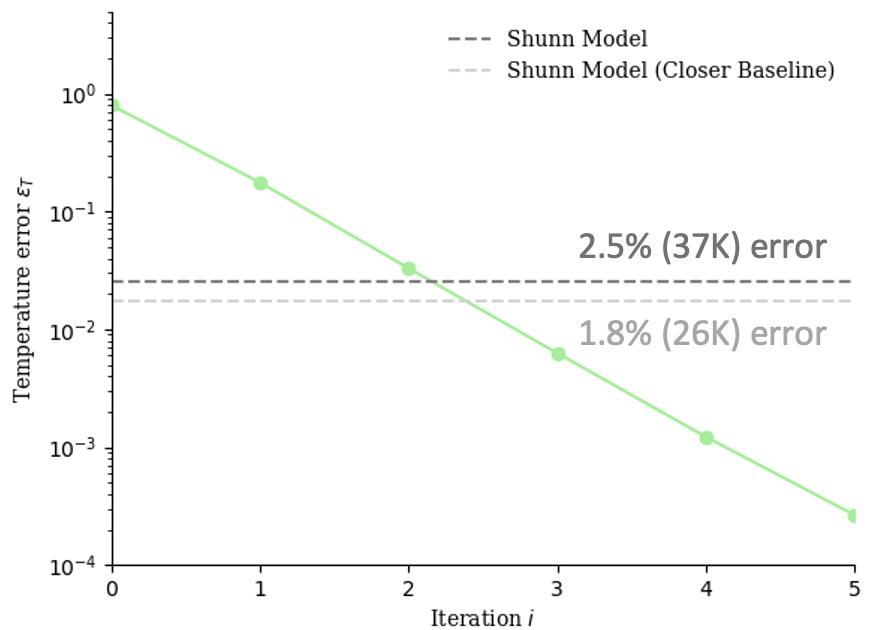}
        \caption{Temperature error}
        \label{fig:strongly-temp-err}
    \end{subfigure}

    \begin{subfigure}[b]{0.49\textwidth}
        \centering
        \includegraphics[width=\textwidth]{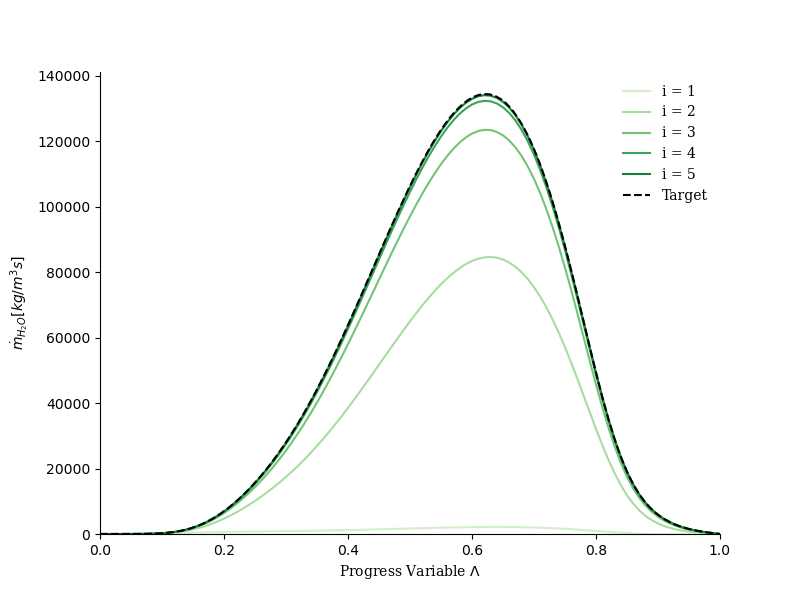}
        \caption{H$_2$O source term iterations}
        \label{fig:strongly-src-iterations}
    \end{subfigure}
    \hfill
    \begin{subfigure}[b]{0.49\textwidth}
        \centering
        \includegraphics[width=\textwidth]{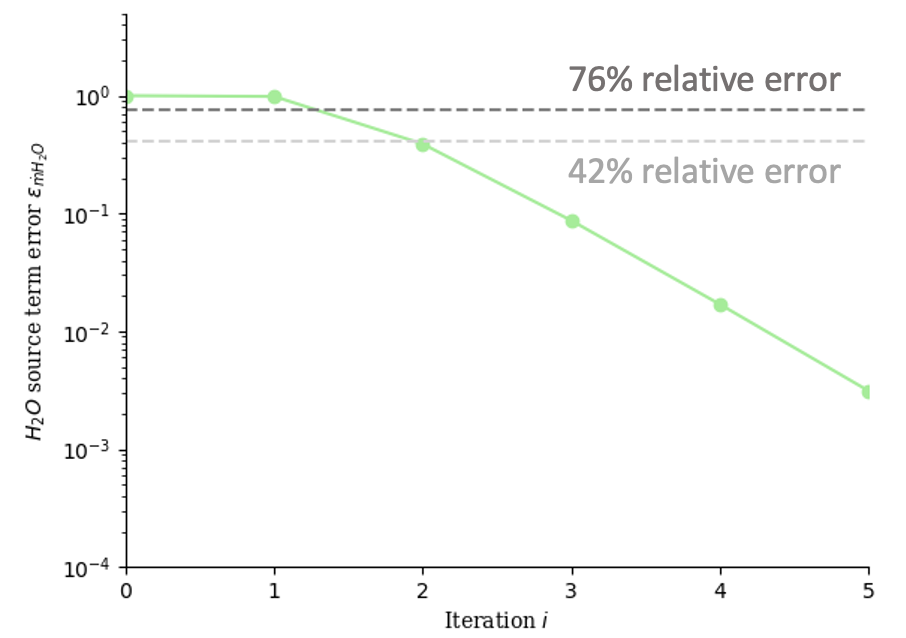}
        \caption{H$_2$O source term error}
        \label{fig:strongly-src-err}
    \end{subfigure}

    \caption{Convergence of the model for the compressed deflagration case. Top row: temperature profiles and error; bottom row: H$_2$O source term profiles and error. Errors compared at $\Lambda=0.5$.}
    \label{fig:strongly-compressed}
\end{figure*}

The thermodynamically consistent model is then applied to the compressed deflagration and the results presented in Fig. \ref{fig:strongly-compressed}. The model was provided $\rho$, $e$, and $\chi_{\Lambda \Lambda}$ at a progress variable of $\Lambda=0.5$ from the compressed 1D deflagration. The model iterates to the correct temperature, as seen in Fig. \ref{fig:strongly-temp-iterations}, reaching sub $1$\% temperature relative errors (see Fig. \ref{fig:strongly-temp-err}) in only three iterations from an initial $\widetilde{RT}$ corresponding to an unburned $T=300$ K and $p=1$ atm. The distant initial guess is used to illustrate the robustness of the model. Fewer iterations are required when a better initial guess is provided; in LES, a good initial guess for $\widetilde{RT}$ is the converged $\widetilde{RT}$ from the previous timestep which would likely require fewer iterations \cite{Yao_2025}. Importantly, the model also captures the correct profile and value for $\dot{m}_{{\rm H}_2{\rm O}}$, as displayed in Figs. \ref{fig:strongly-src-iterations} and \ref{fig:strongly-src-err}.
For compressible deflagrations, a notable benefit of the model is that temperature and source term errors can be minimized arbitrarily with sufficient iterations. Similar results were tested and achieved for expanded cases as shown in Appendix~\ref{sec:additional-one-dimensional-analysis}. Note that the Baumgart \cite{Baumgart} model (tabulated ZND detonations) was not used for comparison against the compressed deflagration of Fig. \ref{fig:strongly-compressed} since it produces very large errors when used for deflagrations.

\subsection{ZND Detonation}
\label{ZND detonation}
The ZND results are presented in Fig. \ref{fig:znd-case}. This time, the algorithm is applied pointwise in $\Lambda$ starting from $\Lambda = 0$ and increasing in increments of $0.1$ until $\Lambda = 1.0$. The resulting discrete solutions are then interpolated to reconstruct the full flame structure. Since the model iterates fixed $h,p$ flames, the profile of the resultant thermochemical state matched the structure of the deflagration in Sec. \ref{Compressed Deflagration}. However, for detonations, this pointwise approach helps illustrate how the model can be applied locally to then reproduce a flame structure which, itself, varies in $h,p$ across $\Lambda$. In fact this approach is similar to how the algorithm would be applied in LES and highlights the ability of the model to produce the correct detonation structure over a range of cells. Figure \ref{fig:znd-case} displays local model convergence at each selected $\Lambda$; the results for temperature and $\dot{m}_{{\rm H}_2{\rm O}}$ are provided in Figs. \ref{fig:znd-temp} and \ref{fig:znd-src} respectively. A noteworthy difference for detonations is that the error cannot be minimized arbitrarily. However, excellent agreement is still achieved. Using a source term error normalized by the maximum reference value (which avoids the denominator vanishing at the ends of the domain) results in a maximum error of $5.5\%$ for the model at $\Lambda=0.7$, compared to an order of magnitude difference for the Shunn \cite{Shunn} model at $\Lambda = 0.1$. Again, the Shunn model is expected to struggle the further the thermodynamic state is from its baseline profile, and this is particularly evident for the ZND. For this reason, Saghafian \cite{Saghafian} style models, such as the Shunn \cite{Shunn} model, are often not recommended for use in detonations; further evidence of this is provided in Ref. \cite{Baumgart_Yao_Blanquart_2025}. Since the Baumgart \cite{Baumgart} model uses tabulated ZND solutions and would provide the exact output for the ZND, it is not shown in Fig. \ref{fig:znd-case}. It will instead be investigated in Sec. \ref{rde-analysis}.

\subsection{Flame Structure versus Thermodynamic State}
\label{flame-structure}
While the quantities at a target progress variable can be matched with small error, the thermochemical state away from the target does not always match the entire profile across $\Lambda$ --- for example, in a detonation which does not have constant $h,p$. In other words, the model does not always predict the entire correct profile when matched to a single progress variable. To understand the impact of this, the influence of flame structure was examined by incorporating the PDF. Beta distributions $\widetilde{P}(\Lambda; \widetilde{\Lambda}, \Lambda_v)$ centered on $\widetilde{\Lambda}=0.5$ were used with normalized variances of $\Lambda_v/\left[\widetilde{\Lambda}(1-\widetilde{\Lambda})\right] = 0.01$, $0.1$, and $0.25$. The filtered temperature $\widetilde{T}$ is then obtained by convolving the temperature profile against the PDF:
\begin{equation}
    \widetilde{T}=\int T(\Lambda;\chi_{\Lambda \Lambda},\widetilde{h},\bar{p})\widetilde{P}(\Lambda;\widetilde{\Lambda}, \Lambda_v)d\Lambda
    \label{eq:temp-convolution}.
\end{equation}

\begin{figure}[!t]
    \centering
    \includegraphics[width=0.48\textwidth]{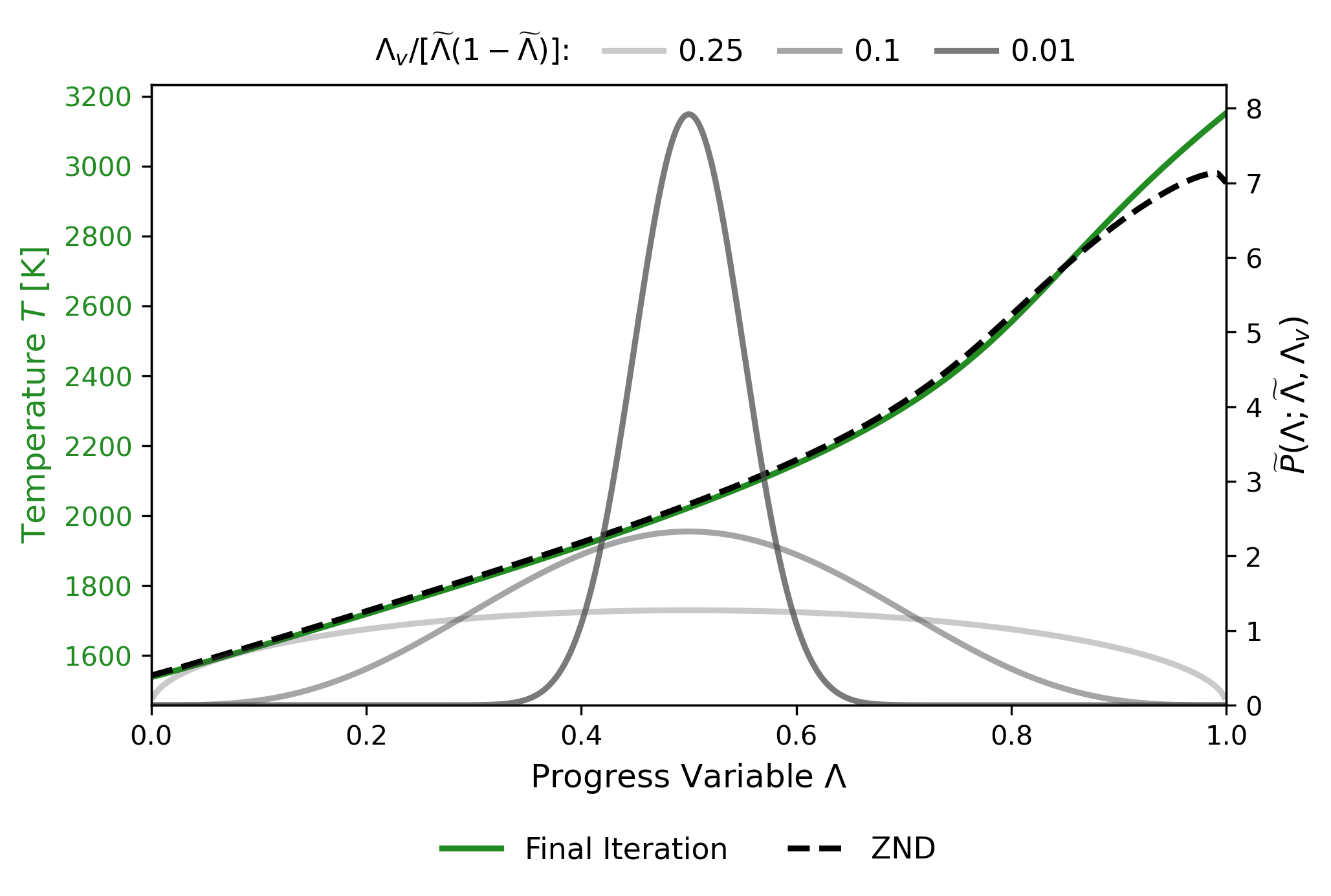}
    \caption{Converged 1D and ZND $T(\Lambda)$ profiles plotted alongside normalized beta PDFs $\widetilde{P}(\Lambda;\widetilde{\Lambda}, \Lambda_v)$ each centered at $\widetilde{\Lambda}=0.5$ with non-dimensional variances of $\Lambda_v/\left[\widetilde{\Lambda}(1-\widetilde{\Lambda})\right] = 0.01$, $0.1$, and $0.25$.
    \label{fig:conv}}
\end{figure}

The profile resulting from convergence at $\Lambda = 0.5$ in Fig. \ref{fig:znd-case} is used alongside the actual ZND $T(\Lambda)$ and $\dot{m}_{{\rm H}_2{\rm O}}(\Lambda)$ profiles, which are then convolved against the various PDFs. Figure \ref{fig:conv} shows the $T(\Lambda)$ profiles with the three normalized beta distributions. Increasing variance captures a greater portion of the $T(\Lambda)$ profile, incorporating more difference in flame structure between the iterated $T(\Lambda)$ profile and the true ZND $T(\Lambda)$ profile.

The results of this analysis are displayed in Table \ref{table:conv}, and it is expected that a larger $\Lambda_v$ will incorporate more of the difference in profile, and subsequently, accrue more error in both temperature and source term. While the results in Table \ref{table:conv} do demonstrate agreement with this expectation, the differences in error are small and mostly saturated. This result stems from the fact that the most notable difference in flame structure occurs far from the matched progress variable of $\Lambda=0.5$ so the variance would have to be very large before that difference could be incorporated.

At this point it is worth reiterating that the Shunn \cite{Shunn} model exhibits temperature errors on the order of several hundred Kelvin \emph{before} any convolution, so these errors are quite small in comparison. Furthermore, the results presented here demonstrate that reproducing the correct background thermodynamic state is the primary requirement for model accuracy, with the flame structure exerting only a secondary influence.

\begin{table}[!t]
    \caption{\label{table:conv}Filtered temperatures $\widetilde{T}$ and relative $\overline{\dot{m}}_{{\rm H}_2{\rm O}}$ error obtained from convolving the converged and ZND profiles against beta distributions with increasing variance $\Lambda_v^\ast = \Lambda_v/\left[\widetilde{\Lambda}(1-\widetilde{\Lambda})\right]$.}
    \centering
    \renewcommand{\arraystretch}{1.4}
    \fontsize{8pt}{9pt}\selectfont
    \setlength{\tabcolsep}{4pt}
    \begin{tabular}{lcccc}
    \hline
    $\Lambda_v^\ast$ & $\widetilde{T}_{ZND}$ [K] & $\widetilde{T}_{model}$ [K] & $\widetilde{T}_{diff}$ [K] & $\overline{\dot{m}}_{{\rm H}_2{\rm O}, err}$ [\%]\\
    \hline
    0.01 & 2034.2 & 2024.3 &   9.9 & 4.62\\
    0.1  & 2060.2 & 2049.5 &   10.7 & 4.89\\
    0.25  & 2106.9 & 2101.7 &  5.2 & 4.00\\
    \hline
    \end{tabular}
\end{table}

\section{A priori RDE Analysis}
\label{rde-analysis}
\begin{figure}[!t]
    \centering
    \includegraphics[width=0.48\textwidth]{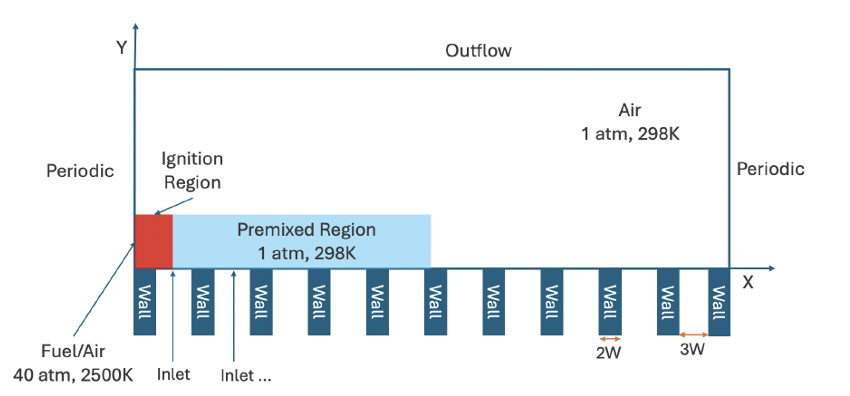}
    \caption{Schematic of the RDE-like computational domain. Ignition, premixed, and ambient air regions are indicated by red, blue, and white shading, respectively.
    \label{fig:rde-setup}}
\end{figure}

Building on the preceding one-dimensional analysis, the model is next evaluated a priori against high-fidelity data from an RDE-like simulation. The simulation was carried out with PeleC \cite{PeleC}, a compressible reacting flow solver built on the adaptive mesh refinement framework AMReX \cite{Amrex}. Unity Lewis numbers are enforced to eliminate differential diffusion effects and focus on the primary modeling challenge. The setup follows previous work \cite{Valencia} as shown schematically in Fig. \ref{fig:rde-setup}. A two-dimensional, unwrapped geometry is employed, with periodic boundary conditions applied at the streamwise ($x$) boundaries. The lower boundary consists of alternating no-slip wall segments and 60 inlet sections. Inlet conditions are prescribed using isentropic nozzle relations with a total pressure of $500$ kPa and a total temperature of $298.15$ K. The inflow state is determined dynamically based on the interior pressure, allowing for subsonic, sonic, or choked flow, and injects a stoichiometric H$_2$-air mixture accordingly. The upper axial ($y$) boundary is treated with a simple outflow condition, with further details provided in Ref. \cite{Valencia}. To initiate the detonation, a thin ignition region (stoichiometric H$_2$-air mixture at $T=2500$ K and $p=40$ atm) is placed immediately downstream of an unburned premixed region (stoichiometric H$_2$-air mixture at $T=298.15$ K and $p=1$ atm); otherwise the domain is filled with air at $T=298.15$ K and $p=1$ atm. 

The domain measures $15$ cm in the streamwise ($x$) direction and $5$ cm in the axial ($y$) direction, discretized with 600 and 200 cells on the base grid, respectively, yielding a base resolution of $250~\mu\mathrm{m}$. Three levels of adaptive mesh refinement (AMR) are applied based on pressure and density gradients, resulting in a finest resolution of $31.25~\mu\mathrm{m}$. An additional refinement criterion based on the OH mass fraction $Y_{\rm OH}$ is included to better resolve deflagrative flame structures. According to the Shock and Detonation Toolbox \cite{edl_sdtoolbox_2023}, the induction length of a stoichiometric CJ H$_2$-air detonation is approximately $190~\mu\mathrm{m}$, indicating that the finest AMR resolution should yield between 5 to 10 grid points within the induction zone. Although this resolution likely does not fully resolve the internal detonation structure, the present RDE simulation is used primarily as a source of representative thermochemical states and quasi-one-dimensional flame structures for model comparison. Initialization effects dissipate after approximately five wave cycles. All data presented here was collected shortly thereafter. Figure \ref{fig:rde-deflag-image} provides an example of the field used for analysis.

The following subsections proceed with an a priori validation of the model using the high-fidelity RDE-like dataset. The performance of the new thermodynamically consistent model is compared against existing approaches to assess its predictive capability. Similar to Sec. \ref{1d-analysis}, the analysis will be performed first against a local deflagration and then a local detonation to demonstrate the versatility of the model. For both, quasi-1D data is obtained by extracting a slice through through the corresponding flame structure along either the x- or y-direction; the resulting profiles are displayed in Figs. \ref{fig:rde-deflag-slice} and \ref{fig:rde-det-slice}. The analysis is performed pointwise (in the same manner as Sec. \ref{ZND detonation}) across the flame structure for both the deflagration and detonation.

\subsection{Model Validation: Deflagration}
\begin{figure*}[!t]
    \centering
    \begin{subfigure}[c]{0.48\textwidth}
        \centering
        \includegraphics[width=\linewidth]{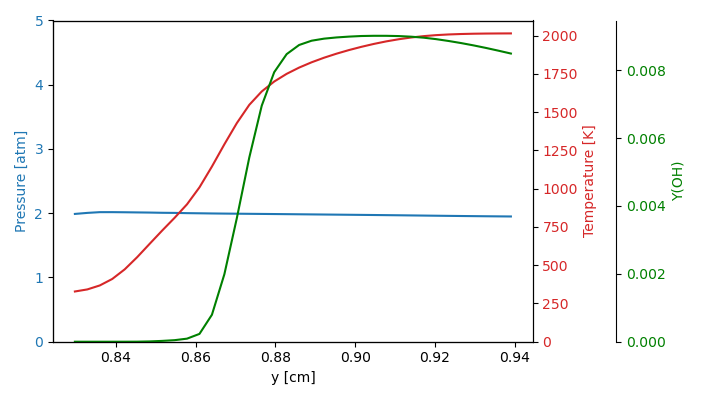}
        \caption{Extracted quasi-1D profiles for pressure, temperature, and OH mass fraction ($Y_{\rm OH}$).}
        \label{fig:rde-deflag-plot}
    \end{subfigure}
    \hfill
    \begin{subfigure}[c]{0.48\textwidth}
        \centering
        \includegraphics[width=\linewidth]{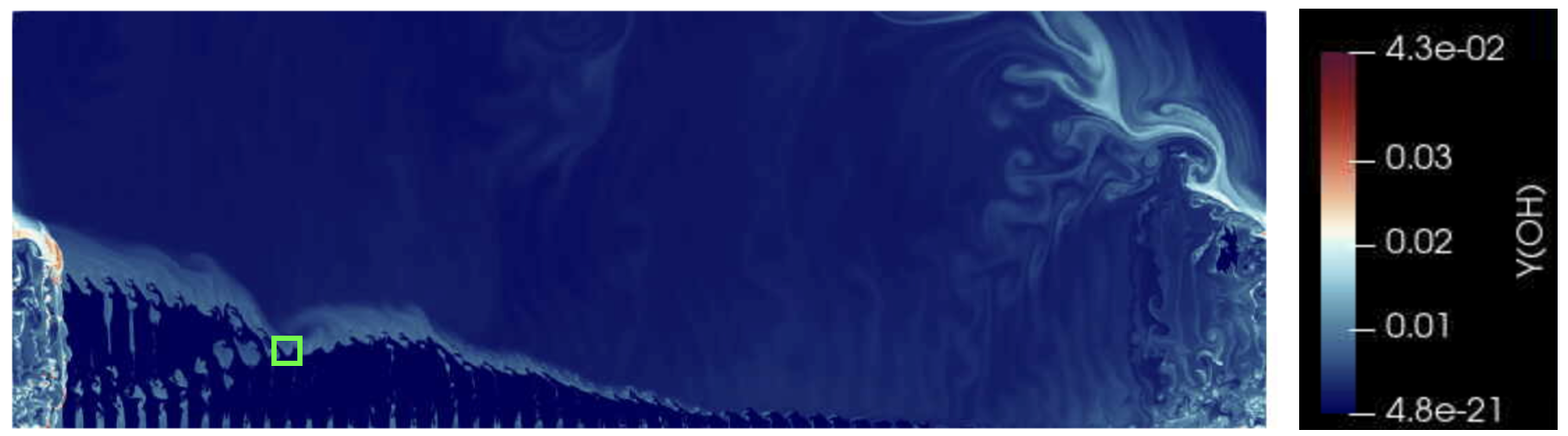}
        \caption{Extraction location in the RDE domain highlighted by the green box.}
        \label{fig:rde-deflag-image}
    \end{subfigure}
    \caption{Quasi-1D slice through a parasitic deflagration region in the RDE-like simulation.}
    \label{fig:rde-deflag-slice}
\end{figure*}

\begin{figure*}[!t]
    \centering
    \begin{subfigure}{0.45\textwidth}
        \centering
        \includegraphics[width=\linewidth]{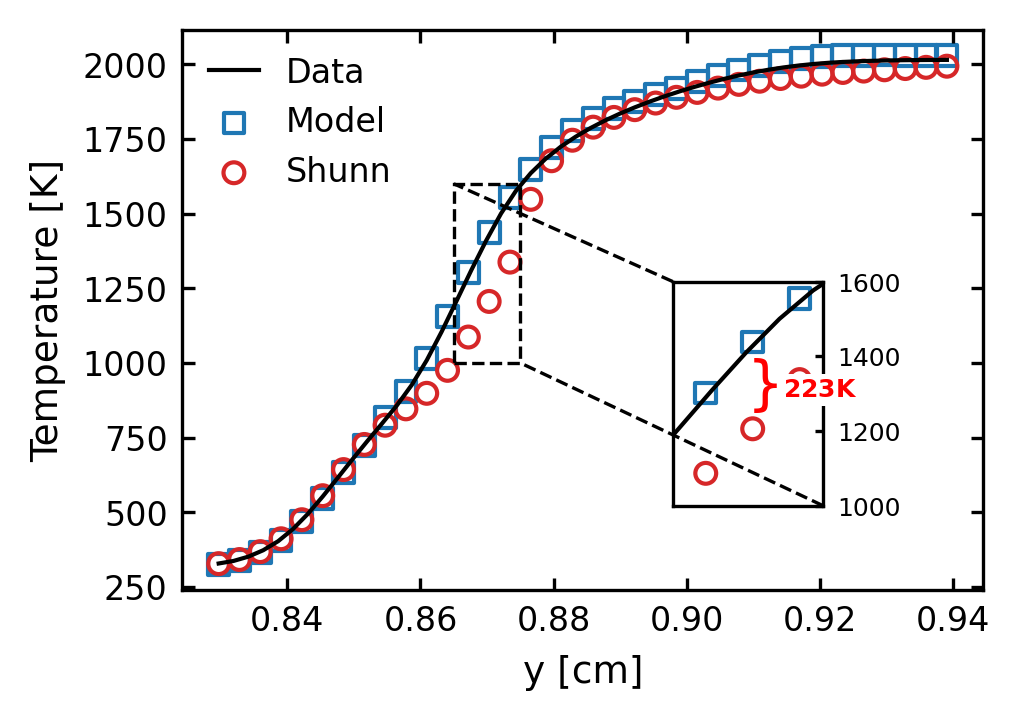}
        \caption{Temperature predictions}
        \label{fig:deflag-temp-compare}
    \end{subfigure}
    \hfill
    \begin{subfigure}{0.45\textwidth}
        \centering
        \includegraphics[width=\linewidth]{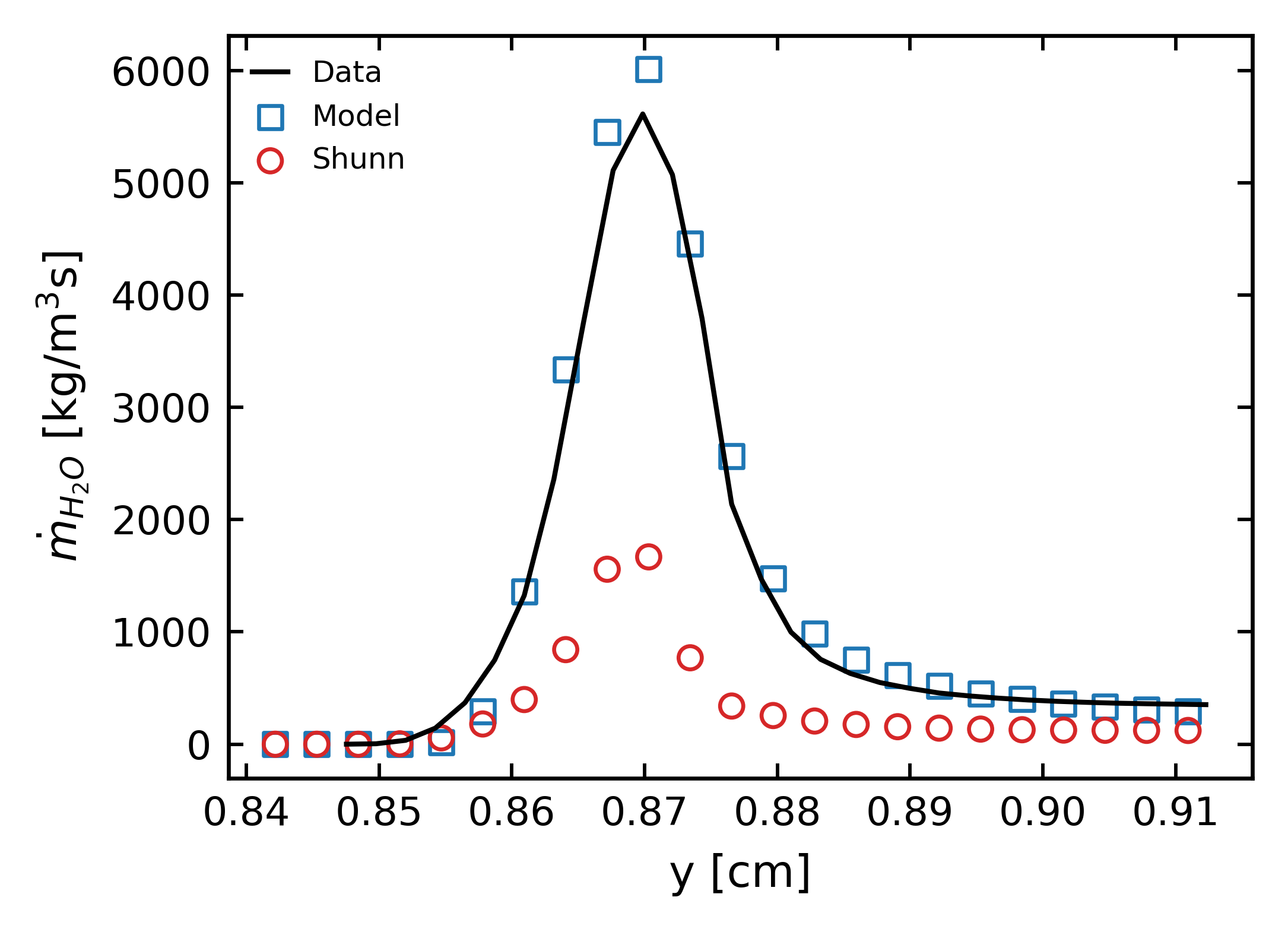}
        \caption{$\dot{m}_{{\rm H}_2{\rm O}}$ predictions}
        \label{fig:deflag-mdot-compare}
    \end{subfigure}
    \caption{Comparison of model predictions against extracted quasi-1D deflagration data in the RDE-like simulation. The thermodynamically consistent model is represented by the blue squares and the Shunn et al. \cite{Shunn} model by the red circles.}
    \label{fig:deflag-model-compare}
\end{figure*}

The deflagration data is displayed in Fig. \ref{fig:rde-deflag-slice} with its extraction location shown. The local $\rho$, $e$, $\chi_{\Lambda \Lambda}$, and $\Lambda$ values were provided to the model and it was assumed that the simulation resolution was sufficient such that the variance was negligible ($\Lambda_{v} \approx 0$). The Shunn model used the baseline deflagration profile presented in Fig. \ref{fig:Shunn-tuning} with the parameters tuned at $\Lambda=0.5$. The iterative algorithm of the thermodynamically consistent model was applied at each point seen in Fig. \ref{fig:deflag-model-compare} until convergence was achieved (generally in 3 to 5 iterations from $T=300$ K and $p=1$ atm). Again, the Baumgart \cite{Baumgart} model is not displayed here since the approach relies on tabulated detonations which produce significant errors when applied to deflagrations.

The results are presented in Fig. \ref{fig:deflag-model-compare}. In particular, Fig. \ref{fig:deflag-temp-compare} shows that the model accurately reproduces the observed temperature profile from the data, while the Shunn model underpredicts temperatures by up to $223$ K near values of $\Lambda=0.5$, despite the tuning parameters being calibrated at this value. This discrepancy arises from the combination of a linear fit for the specific heat ratio, $\gamma$, and a linear perturbation around the baseline internal energy, $e-e_o$, which together cannot capture the nonlinear behavior needed to predict the correct temperature. Ref. \cite{Baumgart_Yao_Blanquart_2025} discusses this in more detail and proposes several solutions to improve the tuning of the model. 
However, as noted previously, even with improvements to the tuning process, these models remain fundamentally unable to fully account for a consistent thermodynamic state between model and CFD. This is evident in Fig. \ref{fig:deflag-mdot-compare}, where the Shunn model underpredicts the peak source term by a $\approx 70\%$ relative error. The thermodynamically consistent model, which fully integrates the thermodynamic state, significantly improves the reproduction of the source term profile, demonstrating its capability to capture the relevant physics more accurately than previous approaches.

\subsection{Model Validation: Detonation}
\begin{figure*}[!t]
    \centering
    \begin{subfigure}[c]{0.48\textwidth}
        \centering
        \includegraphics[width=\linewidth]{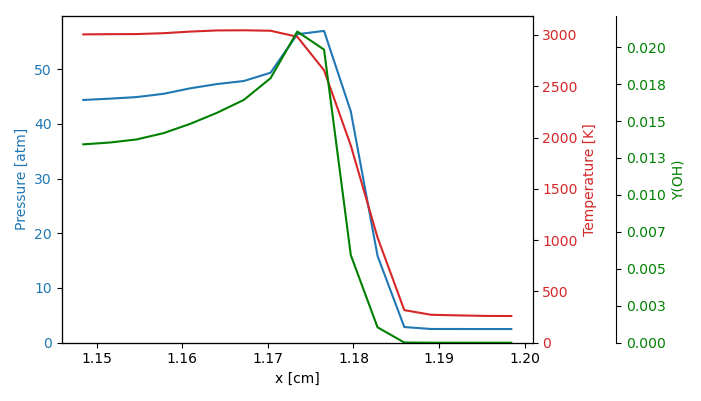}
        \caption{Extracted quasi-1D profiles for pressure, temperature, and OH mass fraction ($Y_{\rm OH}$).}
        \label{fig:rde-det-plot}
    \end{subfigure}
    \hfill
    \begin{subfigure}[c]{0.48\textwidth}
        \centering
        \includegraphics[width=\linewidth]{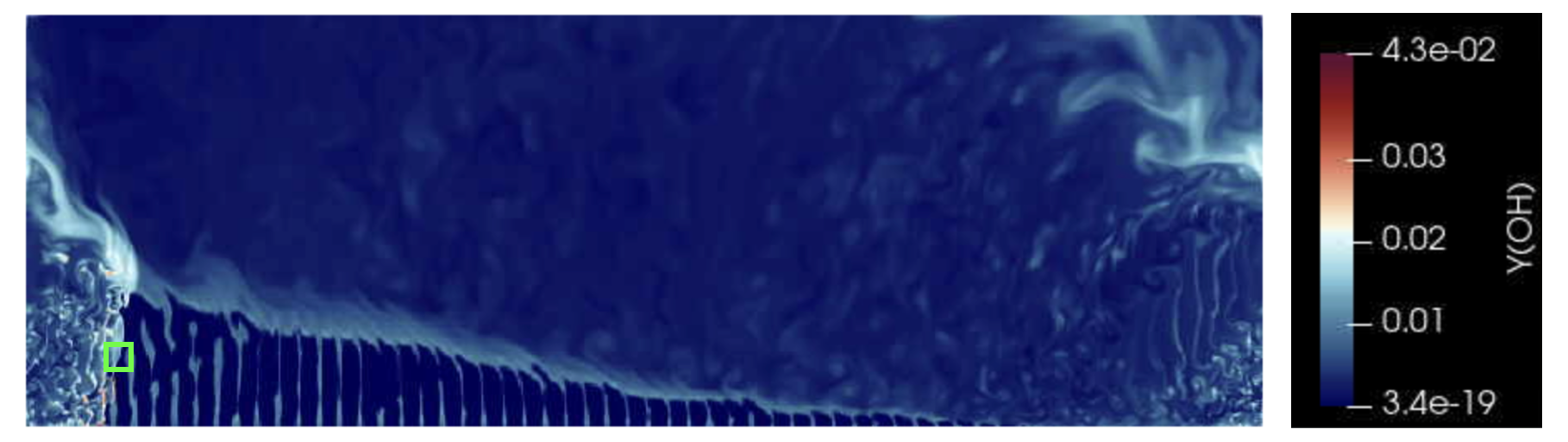}
        \caption{Extraction location in the RDE domain highlighted by the green box.}
        \label{fig:rde-det-image}
    \end{subfigure}
    \caption{Quasi-1D slice through the detonation in the RDE-like simulation.}
    \label{fig:rde-det-slice}
\end{figure*}

\begin{figure*}[!t]
    \centering
    \begin{subfigure}{0.32\textwidth}
        \centering
        \includegraphics[width=\linewidth]{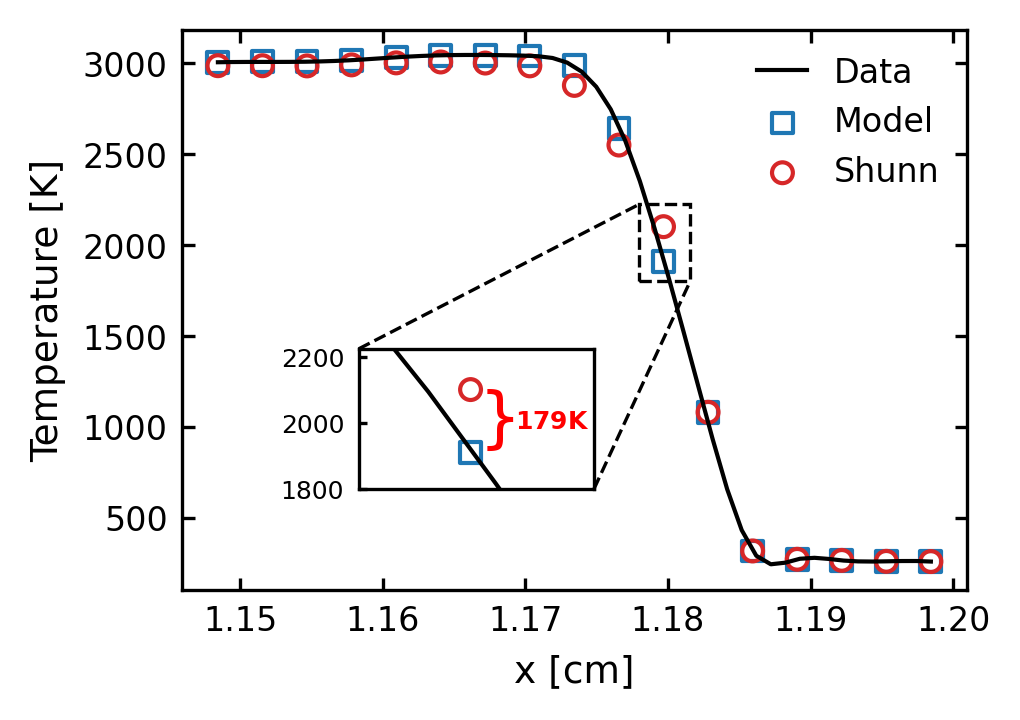}
        \caption{Temperature predictions}
        \label{fig:det-t-compare}
    \end{subfigure}
    \hfill
    \begin{subfigure}{0.32\textwidth}
        \centering
        \includegraphics[width=\linewidth]{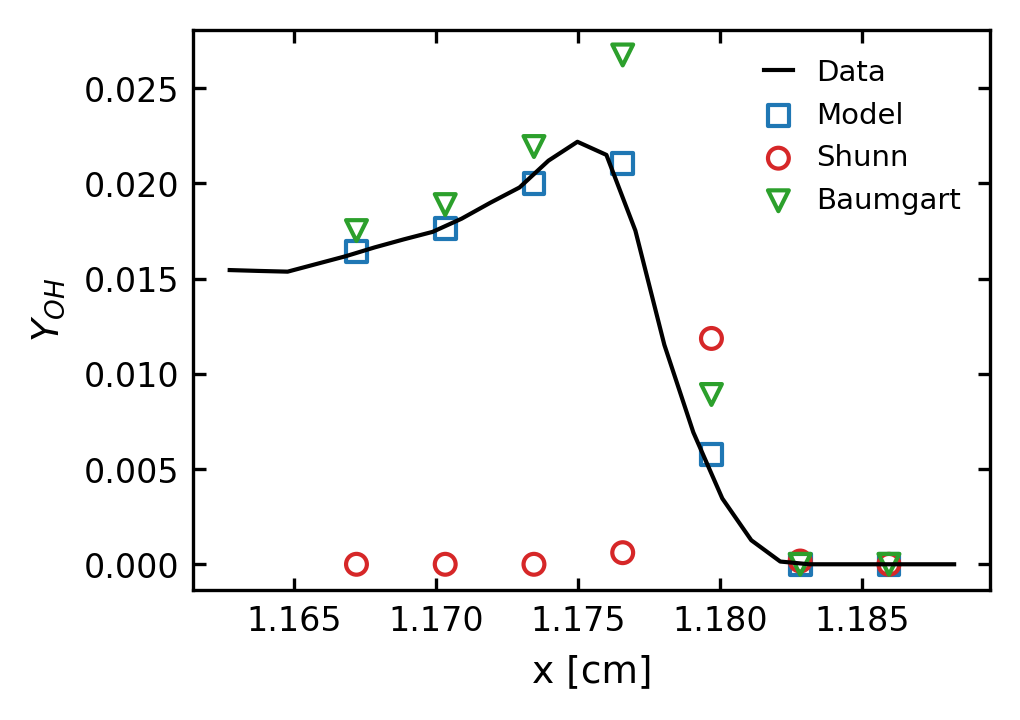}
        \caption{$Y_{\rm OH}$ predictions}
        \label{fig:det-yoh-compare}
    \end{subfigure}
    \hfill
    \begin{subfigure}{0.32\textwidth}
        \centering
        \includegraphics[width=\linewidth]{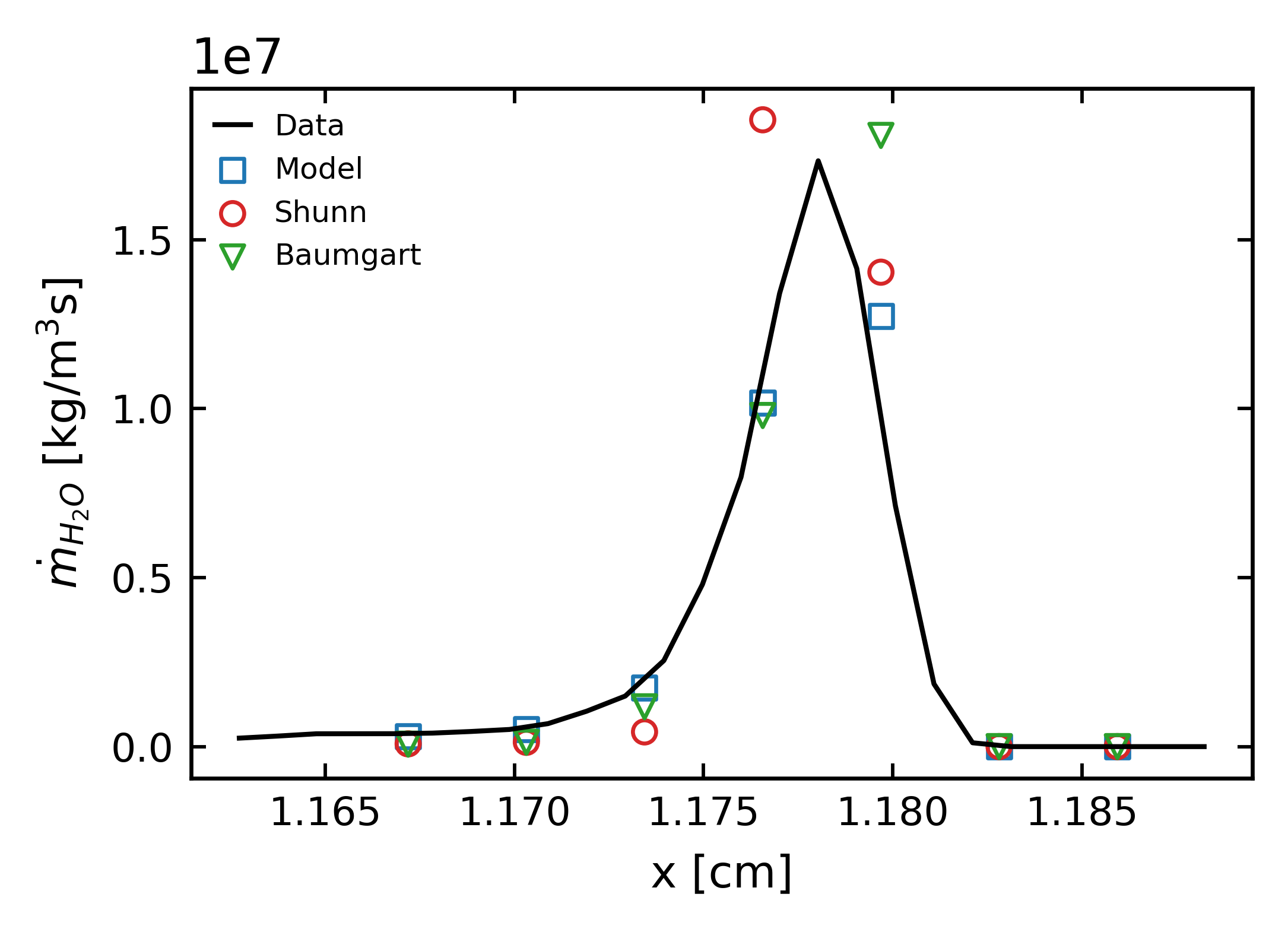}
        \caption{$\dot{m}_{{\rm H}_2{\rm O}}$ predictions}
        \label{fig:det-mdot-compare}
    \end{subfigure}
    \caption{Comparison of model predictions against extracted quasi-1D detonation data in the RDE-like simulation. The thermodynamically consistent model is represented by the blue squares, the Shunn et al. \cite{Shunn} model by the red circles, and the Baumgart et al. \cite{Baumgart} model by the green triangles.}
    \label{fig:det-model-compare}
\end{figure*}

Next, the analysis will examine model performance against a detonation extracted from the data. Models that utilize the Saghafian \cite{Saghafian} framework, such as the Shunn \cite{Shunn} model, are expected to perform poorly for detonations for the reasons mentioned previously. Figure \ref{fig:rde-det-slice}, which shows the temperature, pressure, and OH mass fraction profiles through the detonation, displays a pressure variation of up to 50 atm across the detonation. Deviations of this magnitude from the baseline solution challenge the fundamental premise underlying Saghafian-based \cite{Saghafian} models. Evidence of this is provided in Fig. \ref{fig:det-model-compare}, where the Shunn \cite{Shunn} model overpredicts the temperature by $\approx180$ K (see Fig. \ref{fig:det-t-compare}) in the center of the flame, and subsequently overpredicts the source term with a relative error of $82\%$ at $x\approx1.177$ cm (Fig. \ref{fig:det-mdot-compare}). Additionally, because this model assumes no change in composition from the baseline deflagration-based solution, predictions of species, especially radicals like OH (Fig. \ref{fig:det-yoh-compare}), are far from their true values with errors exceeding an order of magnitude.

\begin{figure*}[!t]
    \centering
    \begin{subfigure}{0.48\textwidth}
        \centering
        \includegraphics[width=\linewidth]{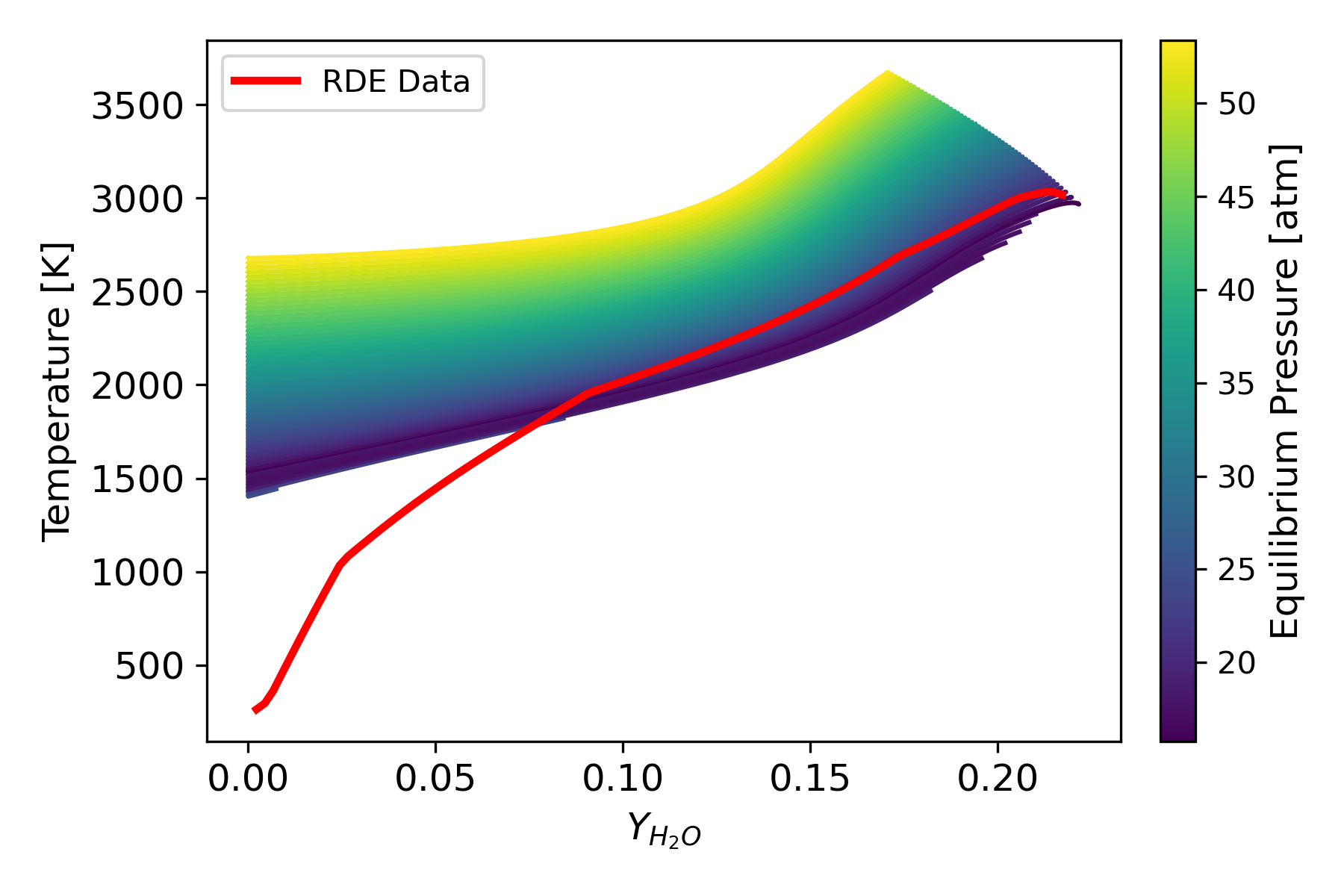}
        \caption{2D tabulation of the ZND profiles with the equilibrium pressure for each profile represented by the color map. The red line shows the corresponding RDE data along the extracted detonation slice.}
        \label{fig:baumgart-compare-2D}
    \end{subfigure}
    \hfill
    \begin{subfigure}{0.48\textwidth}
        \centering
        \includegraphics[width=\linewidth]{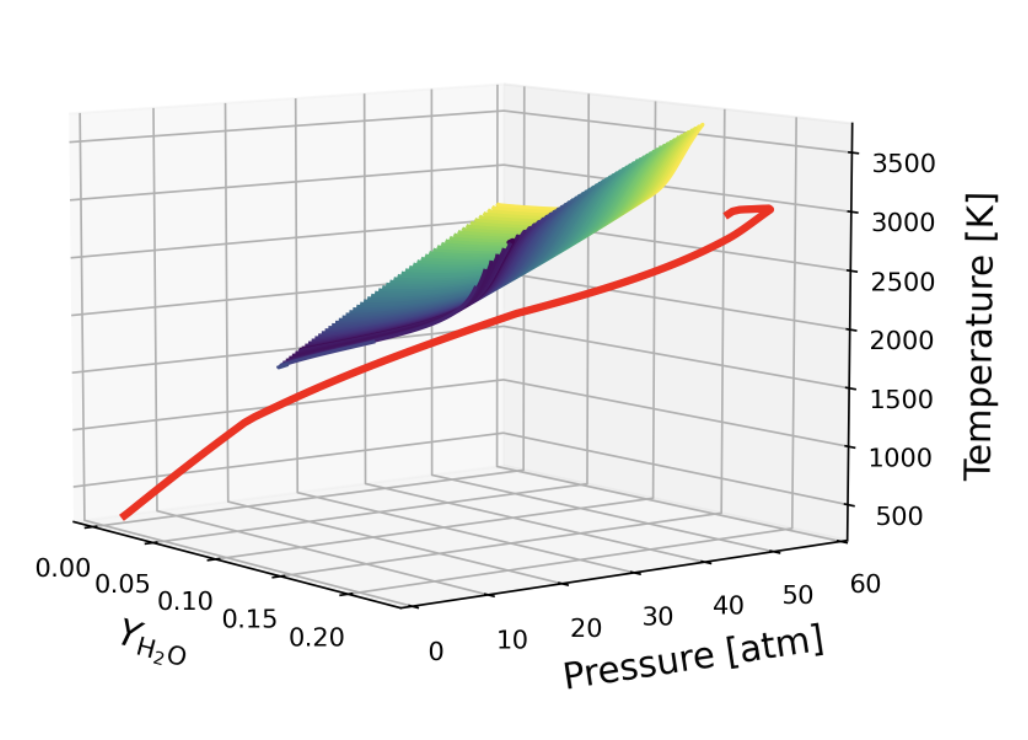}
        \caption{3D representation of the same ZND profiles, with pressure as the third dimension. This visualization highlights the limited overlap between the tabulated profiles and the actual RDE data in 3D space.}
        \label{fig:baumgart-compare-3D}
    \end{subfigure}
    \caption{Visualization of the tabulated ZND detonation profiles used in the Baumgart et al. \cite{Baumgart} model and comparison to the detonation extracted from the RDE-like data.}
    \label{fig:baumgart-compare}
\end{figure*}

Baumgart et al. \cite{Baumgart} addressed this modeling difficulty by changing the underlying flame structure to a detonation. They tabulate a CJ detonation, and the corresponding under- and over-driven ZND detonations, as a function of the conventional progress variable and temperature. Their progress variable source term correction assumes second-order kinetics. This model was implemented here using CJ and under- and over-driven stoichiometric H$_2$-air mixtures with an initial state of $300$ K and $1$ atm.

Since the exact temperature was provided to the Baumgart \cite{Baumgart} model in this work, its performance is only compared against the OH mass fractions and H$_2$O source terms in Fig. \ref{fig:det-model-compare}. The model overpredicts $Y_{\rm OH}$ with a relative error of $32$\%  at $x\approx1.177 $ cm and $\dot{m}_{{\rm H}_2{\rm O}}$ with a relative error of $81$\% at $x\approx1.180$ cm. To understand this, Fig. \ref{fig:baumgart-compare-2D} displays the extracted detonation profile against the tabulated solutions in $(Y_{\rm H_2 O}, ~T)$ space. Before the shock front, the data (represented by the red line) is below the minimum temperatures of the tabulated solutions; however, following the shock and beginning of reaction, the profile overlaps and extracts from the tabulated ZND solutions. Because the tabulated profiles correspond to ZND detonations with a fixed upstream state, the resulting pressure trajectories are also fixed. This is better illustrated in 3D, where pressure is included as an additional dimension (see Fig. \ref{fig:baumgart-compare-3D}).
In this space, it is clear that the pressure of the extracted detonation is lower than that of the tabulated solutions. This discrepancy is accompanied by an overprediction of $Y_{\rm OH}$ and $\dot{m}_{{\rm H}_2{\rm O}}$, visible in Figs. \ref{fig:det-yoh-compare} and \ref{fig:det-mdot-compare}. While one might expect the higher pressure tabulated solutions to exhibit lower radical mass fractions due to enhanced recombination, the opposite trend is observed. A possible explanation is that the assumption of purely bimolecular kinetics interferes with the pressure dependence of the chemistry and the associated radical pool; however, further investigation would be required to determine the exact cause of this behavior. Regardless, the point remains that this approach cannot account for varying thermodynamic states caused by upstream compression, expansion, reshocks, or similar phenomena that ultimately change the behavior of the one-dimensional detonation.

Contrast this with the thermodynamically consistent model that iterates and converges to the correct background thermodynamic state and is subsequently able to capture all the needed information for the thermochemical state, even if the flame structure varies away from the matched progress variable. The success of this approach is demonstrated in Fig. \ref{fig:det-model-compare} where the thermodynamically consistent model best reproduces the observed data compared to the other models which only partially couple the state. To further investigate the role of the subfilter progress variable variance on model error, an analysis similar to that in Sec. \ref{flame-structure} was performed on the RDE data. The results support the conclusions of Sec. \ref{flame-structure}: the model accrues error in the source term with increasing variance (up to $20\%$). However, this is still less than the $\approx80\%$ errors with the other models without considering filtering. This analysis is presented in more detail in Appendix~\ref{sec:variance-analysis}.

As a final point, the model achieves this level of agreement using a fixed h,p flame with a consistent thermodynamic state. The results consistently indicate that model accuracy is governed primarily by the ability to reproduce the local thermodynamic state. In contrast, differences in the underlying flame structure play a comparatively minor role, highlighting the importance of thermodynamic consistency as the primary modeling consideration.

\section{Conclusion}
\label{conclusion}
This work presents the development and validation of a manifold-based turbulent combustion model for compressible and supersonic premixed flames that finds the consistent thermodynamic state between the flow solver and model through an iterative procedure. Validation against one-dimensional and high-fidelity RDE-like data demonstrates that the model quickly converges to the correct thermodynamic state and, as a result, reproduces critical quantities such as temperature, radical species, and source terms, outperforming existing approaches that rely on perturbations around baseline solutions with a nominal thermodynamic state or tabulated detonations with partial-coupling.  This demonstrates the critical importance of consistently accounting for the thermodynamic state, which is more important than any assumptions on the underlying reacting front structure.

The results also illustrate the versatility and robustness of this approach. Unlike prior models, which struggle or fail outside of their respective regimes, this framework provides consistent predictive performance across the full spectrum of compressible flames. However, a posteriori analysis following full implementation and coupling to a flow solver is still needed, particularly to determine how any errors in the thermochemical state influence future states. Such an analysis would also quantify the model's performance and reduction in computational cost. By enabling accurate representation of both deflagrative and detonative processes, the model offers a unified tool for predictive LES of high-speed reacting flows and lays the groundwork for future studies in compressible combustion, including applications to rotating detonation engines and other supersonic reacting systems.

\section*{Acknowledgments}
The authors gratefully acknowledge funding from the Air Force Office of Scientific Research (AFOSR) Award FA9550-21-1-0060 as well as support from Princeton Research Computing and the Princeton MAE Second Year Fellowship.

\clearpage

\bibliographystyle{unsrt}
\bibliography{refs}

\clearpage

\appendix
\raggedbottom

\section{Additional One-Dimensional Analysis}
\label{sec:additional-one-dimensional-analysis}

This section will present additional one-dimensional analysis not presented in the manuscript. This includes an expanded case and a weakly-compressed case, as well as additional analysis on the one-dimensional ZND case. Similar to the compressed case presented in Sec. \ref{1d-analysis}, the unburned states for both the expanded and weakly-compressed cases were obtained by isentropically compressing or expanding a stoichiometric H$_2$-air mixture from $T=300$ K and $p = 1$ atm; these states are provided in Table \ref{table:unburned-sm}. The unburned state for the ZND is the Neumann state ($T\approx1540$ K and $p\approx27.9$ atm) resulting from the CJ detonation of a stoichiometric H$_2$-air mixture with $T=300$ K and $p = 1$ atm.

\begin{table}[!htbp]
  \centering
  \caption{Unburned state for the one-dimensional cases: (a)~expanded
    deflagration and (b)~weakly compressed deflagration. Each case uses a
    stoichiometric H$_2$–air mixture ($\phi = 1$).}
  \label{table:unburned-sm}
  \small
  \begin{tabular}{lcccccc}
    \toprule
    \textbf{Case} & $T$ (K) & $P$ (kPa) & $Y_{\mathrm{H_2}}$ &
    $Y_{\mathrm{O_2}}$ & $Y_{\mathrm{N_2}}$ \\
    \midrule
    (a) & 262 & 63.5  & 0.029 & 0.226 & 0.745 \\
    (b) & 343 & 162 & 0.029 & 0.226 & 0.745 \\
    \bottomrule
  \end{tabular}
\end{table}

\subsection{Expanded Deflagration}
\label{sec:expanded}
\begin{figure}[!htbp]
  \centering
  \begin{subfigure}[b]{0.4\textwidth}
    \centering
    \includegraphics[width=\textwidth]{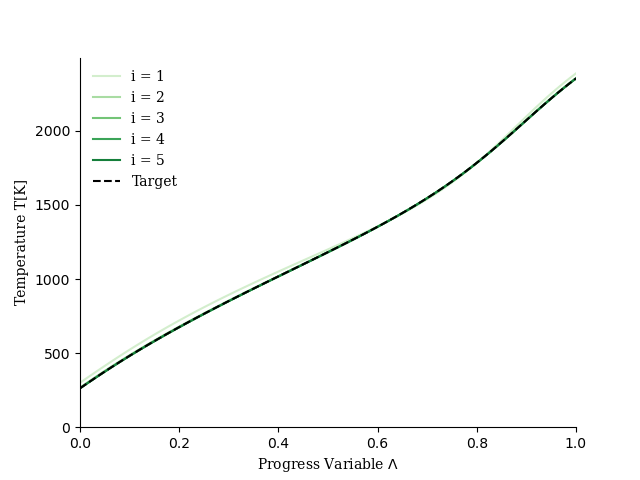}
    \caption{Temperature iterations.}
    \label{fig:expanded-temp-iterations}
  \end{subfigure}
  \hspace{0.04\textwidth}
  \begin{subfigure}[b]{0.4\textwidth}
    \centering
    \includegraphics[width=\textwidth]{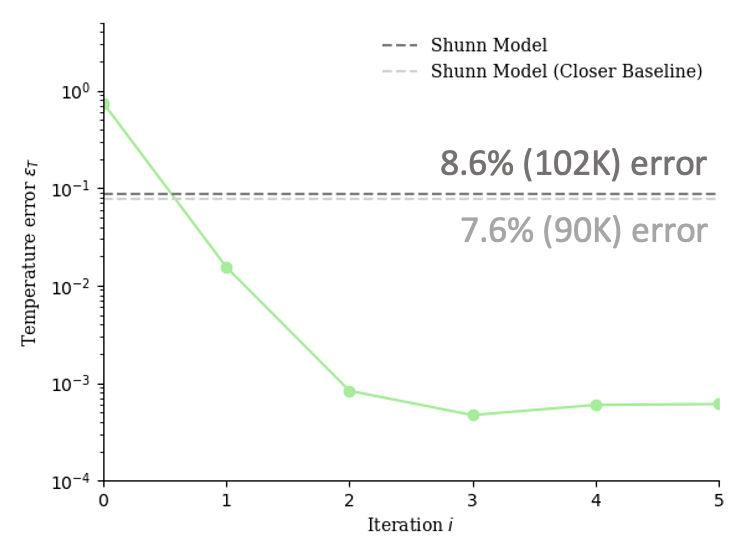}
    \caption{Temperature error.}
    \label{fig:expanded-temp-err}
  \end{subfigure}

  \vspace{8pt}

  \begin{subfigure}[b]{0.4\textwidth}
    \centering
    \includegraphics[width=\textwidth]{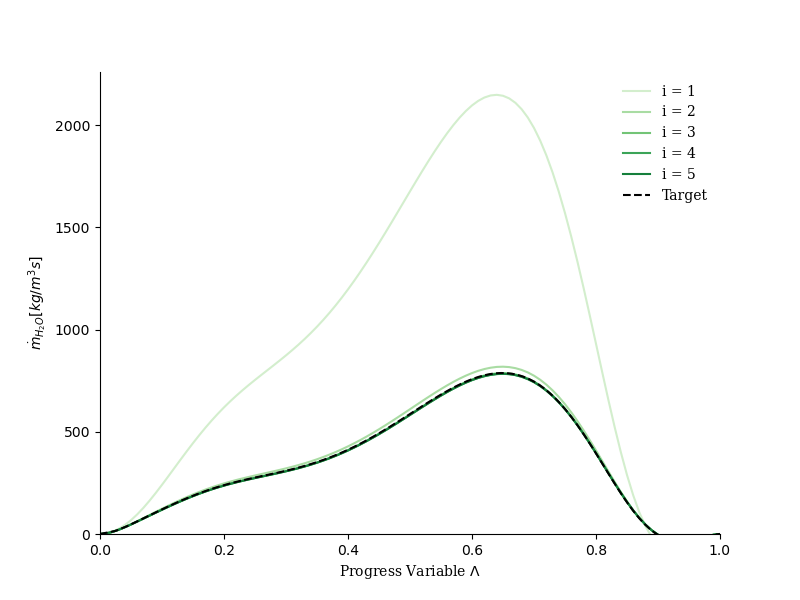}
    \caption{H$_2$O source term iterations.}
    \label{fig:expanded-src-iterations}
  \end{subfigure}
  \hspace{0.04\textwidth}
  \begin{subfigure}[b]{0.4\textwidth}
    \centering
    \includegraphics[width=\textwidth]{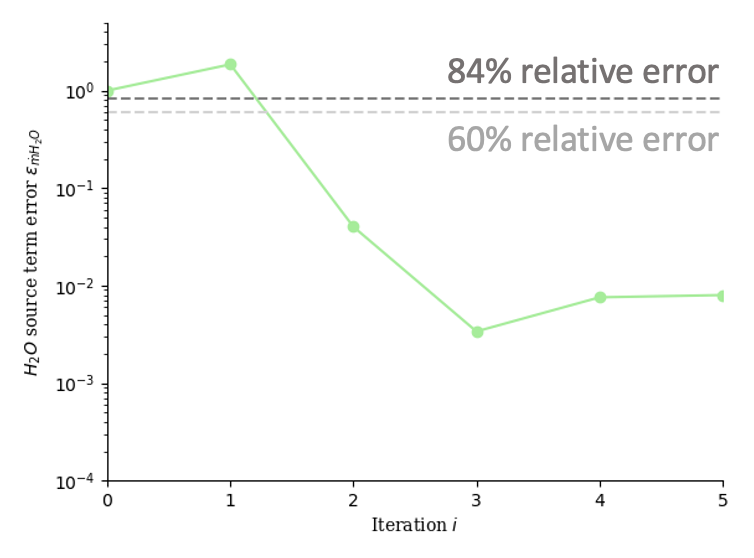}
    \caption{H$_2$O source term error.}
    \label{fig:expanded-src-err}
  \end{subfigure}

  \caption{Convergence of the model for the expanded deflagration case.
    Top row: temperature profiles and error; bottom row: H$_2$O source term
    profiles and error. Errors compared at $\Lambda = 0.5$.}
  \label{fig:expanded}
\end{figure}

The results for the expanded deflagration are presented in Fig.~\ref{fig:expanded}. The thermodynamic state was matched at $\Lambda = 0.5$ and low errors achieved in several iterations. As mentioned previously, for compressible deflagrations the thermodynamically consistent model can achieve arbitrarily small errors provided sufficient iterations. The temperature profiles quickly converge to the target profile (see Fig. \ref{fig:expanded-temp-iterations}), and approximately 0.1\% relative errors are achieved in only two iterations.

The model particularly excels in predicting the water source term $\dot{m}_{\rm H_2 O}$ --- evidence of which is provided in Figs. \ref{fig:expanded-src-iterations} and \ref{fig:expanded-src-err}. By the third iteration sub 1\% relative error is achieved. The results for the Shunn \cite{Shunn} model presented here continue to utilize parameters tuned for the compressed deflagration from the manuscript (with the unburned state described in Table \ref{table:unburned}) from the baseline and closer baseline profiles displayed in Fig. \ref{fig:Shunn-tuning}, and, as a result, produce large relative errors up to $85$\% for the source term, illustrating the lack of robustness of this type of model.

\subsection{Weakly Compressed Deflagration}
\label{sec:weakly-compressed}
\begin{figure}[!htbp]
  \centering
  \begin{subfigure}[b]{0.4\textwidth}
    \centering
    \includegraphics[width=\textwidth]{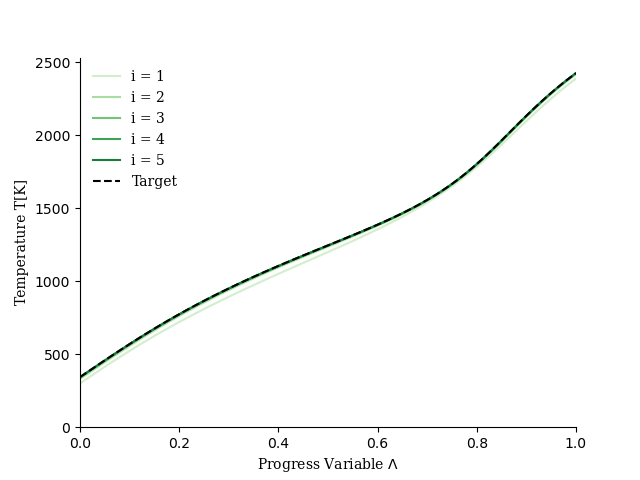}
    \caption{Temperature iterations.}
    \label{fig:weakly-temp-iterations}
  \end{subfigure}
  \hspace{0.04\textwidth}
  \begin{subfigure}[b]{0.4\textwidth}
    \centering
    \includegraphics[width=\textwidth]{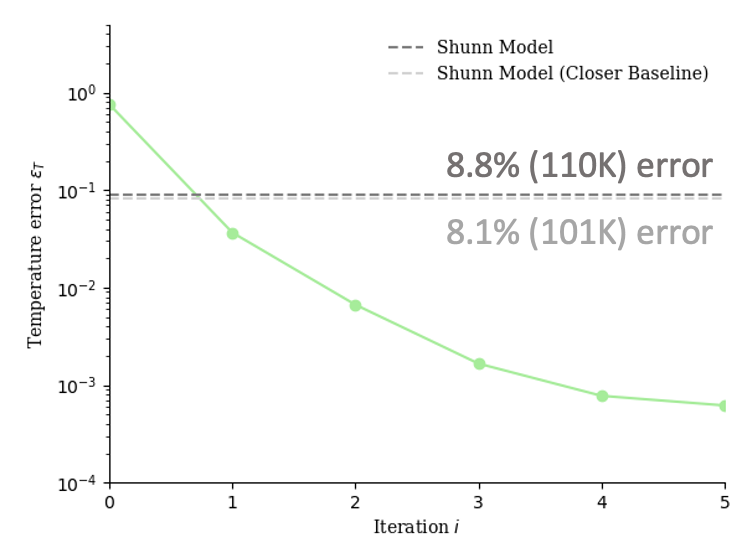}
    \caption{Temperature error.}
    \label{fig:weakly-temp-err}
  \end{subfigure}

  \vspace{8pt}

  \begin{subfigure}[b]{0.4\textwidth}
    \centering
    \includegraphics[width=\textwidth]{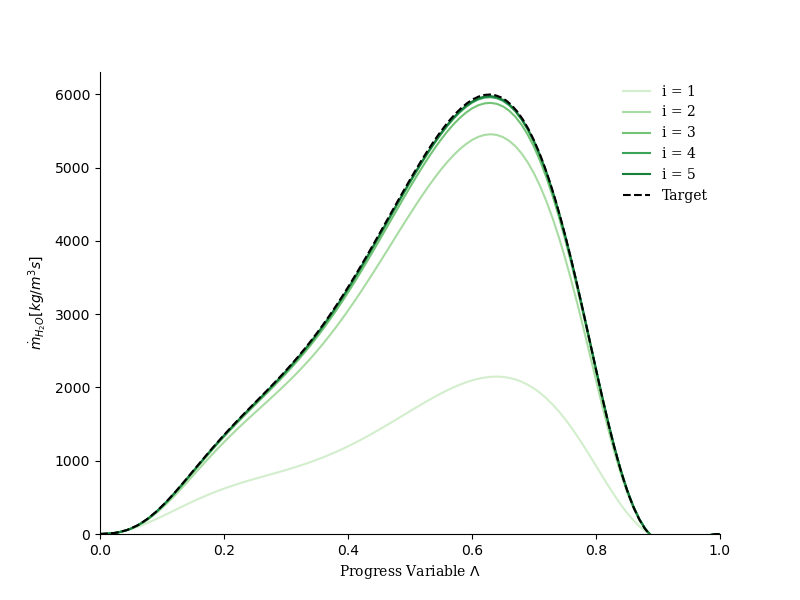}
    \caption{H$_2$O source term iterations.}
    \label{fig:weakly-src-iterations}
  \end{subfigure}
  \hspace{0.04\textwidth}
  \begin{subfigure}[b]{0.4\textwidth}
    \centering
    \includegraphics[width=\textwidth]{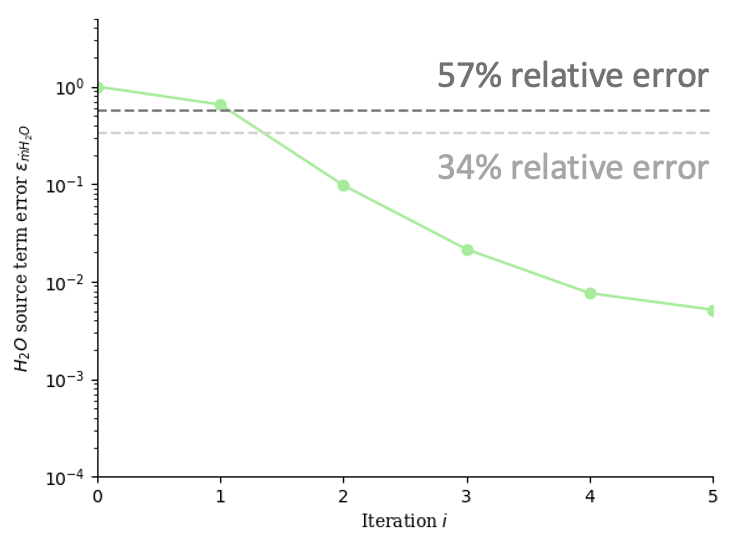}
    \caption{H$_2$O source term error.}
    \label{fig:weakly-src-err}
  \end{subfigure}

  \caption{Convergence of the model for the weakly compressed deflagration case.
    Top row: temperature profiles and error; bottom row: H$_2$O source term
    profiles and error. Errors compared at $\Lambda = 0.5$.}
  \label{fig:weakly-compressed}
\end{figure}

Next, the results for the weakly compressed deflagration are presented in Fig. \ref{fig:weakly-compressed}. The target profile is that associated with the unburned state of (b) in Table \ref{table:unburned-sm}, and again convergence is achieved at $\Lambda = 0.5$. The results provide further evidence to the previously described conclusions. Figure \ref{fig:weakly-temp-iterations} shows quick convergence to the target temperature profile and sub 1\% temperature error is achieved by the second iteration as shown in Fig. \ref{fig:weakly-temp-err}, while the Shunn \cite{Shunn} model produces errors in excess of $100$ K.

Similarly, the $\dot{m}_{\rm H_2 O}$ source term converges quickly (see Fig. \ref{fig:weakly-src-iterations}) and errors can be made arbitrarily small and reach approximately 1\% (Fig. \ref{fig:weakly-src-err}) in four iterations. The Shunn \cite{Shunn} model similarly struggles for the source term, producing a $57$\% relative error for the baseline case, and reinforcing the need for thermodynamic consistency.

\subsection{Detonation}
\label{sec:detonation}

\begin{figure}[!htbp]
  \centering
  \begin{subfigure}[b]{0.4\textwidth}
    \centering
    \includegraphics[width=\textwidth]{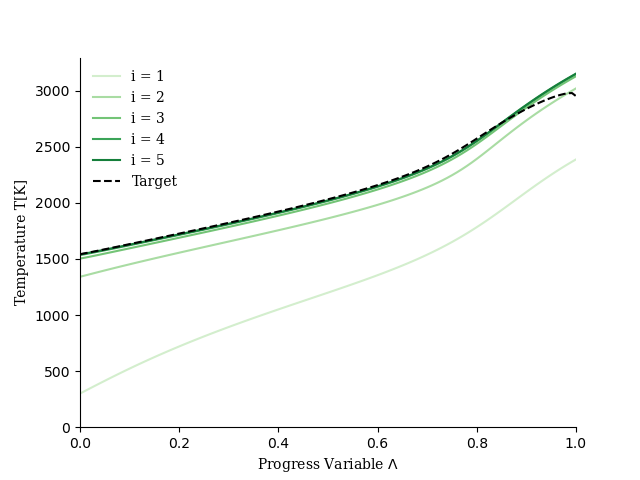}
    \caption{Temperature iterations.}
    \label{fig:znd-temp-iterations}
  \end{subfigure}
  \hspace{0.04\textwidth}
  \begin{subfigure}[b]{0.4\textwidth}
    \centering
    \includegraphics[width=\textwidth]{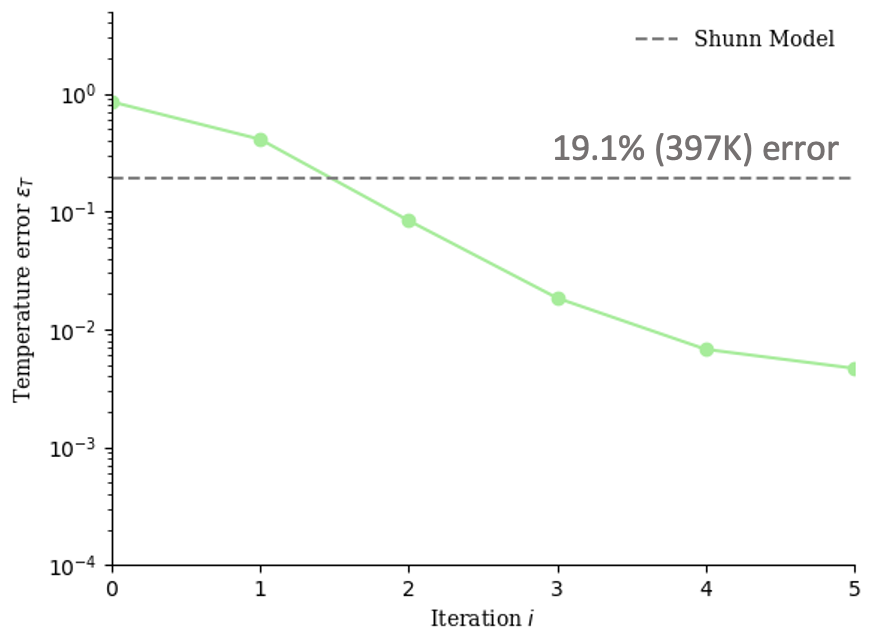}
    \caption{Temperature error.}
    \label{fig:znd-temp-err}
  \end{subfigure}

  \vspace{8pt}

  \begin{subfigure}[b]{0.4\textwidth}
    \centering
    \includegraphics[width=\textwidth]{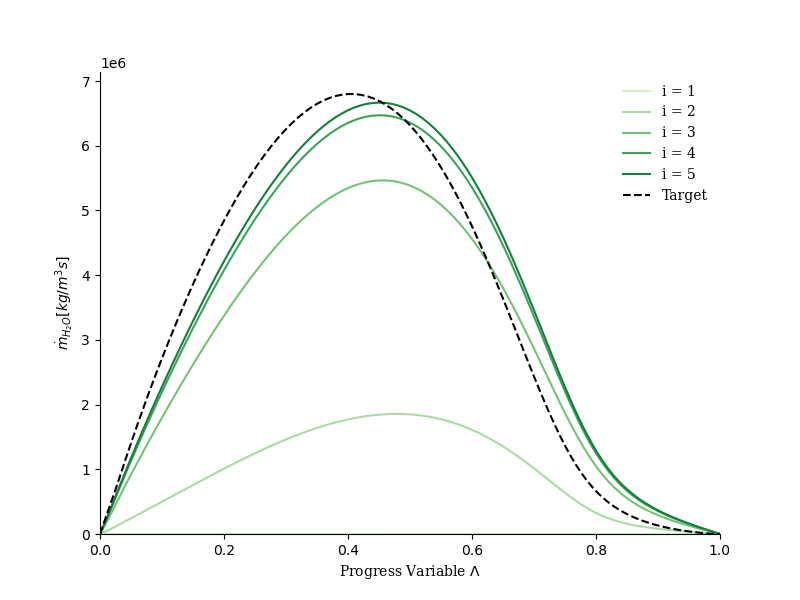}
    \caption{H$_2$O source term iterations.}
    \label{fig:znd-src-iterations}
  \end{subfigure}
  \hspace{0.04\textwidth}
  \begin{subfigure}[b]{0.4\textwidth}
    \centering
    \includegraphics[width=\textwidth]{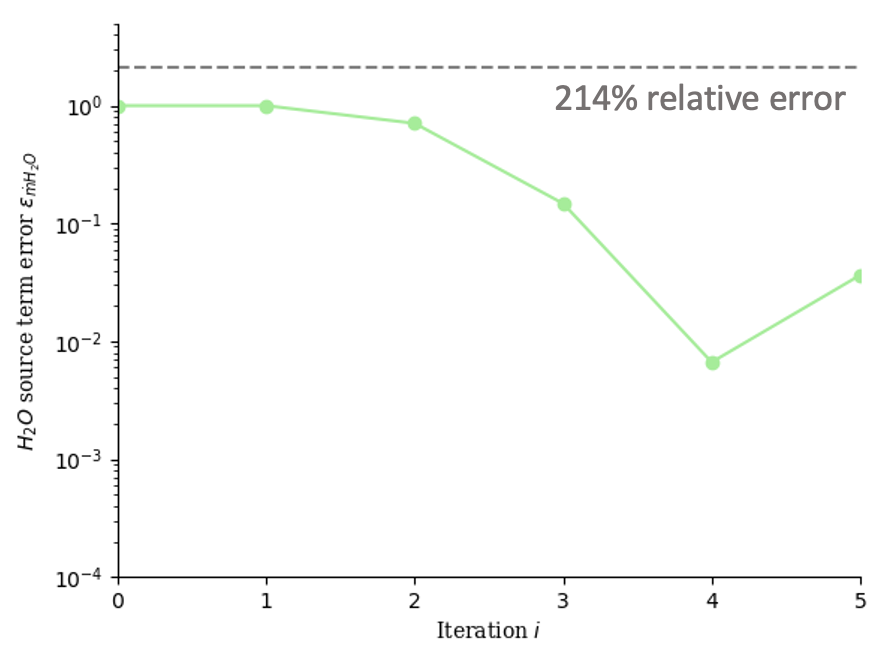}
    \caption{H$_2$O source term error.}
    \label{fig:znd-src-err}
  \end{subfigure}
  \caption{Convergence of the model for the one-dimensional detonation case.
    Top row: temperature profiles and error; bottom row: H$_2$O source term
    profiles and error. Errors compared at $\Lambda = 0.5$.}
  \label{fig:znd}
\end{figure}

The manuscript presented an alternative analysis for the one-dimensional detonation case which demonstrated convergence at various $\Lambda$ to illustrate the accuracy of the model over the $T(\Lambda)$ and $\dot{m}_{\rm H_2 O}(\Lambda)$ profiles. For completion, this section additionally provides the familiar analysis of $T$ and $\dot{m}_{\rm H_2 O}$ convergence at $\Lambda = 0.5$. These results are presented in Fig. \ref{fig:znd}. Notably, detonations do not have constant $h,p$ and, as such, possess a different flame structure from that which is iterated by the algorithm. The differences in the iterated $T(\Lambda)$ and $\dot{m}_{\rm H_2 O}(\Lambda)$ from the actual ZND $T(\Lambda)$ and $\dot{m}_{\rm H_2 O}(\Lambda)$ are evident in Figs. \ref{fig:znd-temp-iterations} and \ref{fig:znd-src-iterations}. This is better illustrated in Fig. \ref{fig:znd-src-iterations} where the peak source term occurs earlier in the ZND compared to the constant $h,p$ flame. As a result of this difference in flame structure, the errors cannot be made arbitrarily small like the deflagration cases.

With that in mind, the important criterion for accuracy is a matched thermodynamic state at the given progress variable ($\Lambda =0.5$ in this case). This cannot be perfectly achieved due to the difference in flame structure but $\approx 1$\% errors are still possible for both temperature (see Fig. \ref{fig:znd-temp-err}) and source term (Fig. \ref{fig:znd-src-err}). Contrast this with the Shunn \cite{Shunn} model which yields a $\approx 400$ K error and $214\%$ source term relative error. Furthermore, Sec. \ref{ZND detonation} in the manuscript demonstrates that errors resulting from convolution against a differing flame structure are considerably less important than matching the local thermodynamic state.

\FloatBarrier
\section{Progress Variable Definition Analysis}
\label{sec:source-term}

The compressible progress variable source term derived in Sec. \ref{sec:transport} is:
\begin{equation}
  \dot{m}_{\Lambda_c} =
    \frac{\rho\,\chi_{\Lambda Y_{R,\mathrm{eq}}} + \dot{m}_{R}
          - \Lambda\,\dot{m}_{Y_{R,\mathrm{eq}}}}{Y_{R,\mathrm{eq}} - Y_{R,u}},
  \label{eq:high-speed-source-term-sm}
\end{equation}
where
\begin{equation}
  \chi_{\Lambda Y_{R,\mathrm{eq}}} =
    2D\,\frac{\partial\Lambda}{\partial x_j}
    \frac{\partial Y_{R,\mathrm{eq}}}{\partial x_j}
  \label{eq:chi-ly-sm}
\end{equation}
and
\begin{equation}
  \dot{m}_{Y_{R,\mathrm{eq}}} =
    \frac{\partial\rho Y_{R,\mathrm{eq}}}{\partial t}
    + \frac{\partial\rho u_j Y_{R,\mathrm{eq}}}{\partial x_j}
    - \frac{\partial}{\partial x_j}
      \!\left(\rho D\,\frac{\partial Y_{R,\mathrm{eq}}}{\partial x_j}\right).
  \label{eq:source-term-y-ref-eq-sm}
\end{equation}

The high-fidelity RDE-like dataset presented in the manuscript is leveraged to address the remaining unclosed terms in $\dot{m}_{\Lambda_c}$ (Eq. \ref{eq:high-speed-source-term-sm}). These include the coupling scalar dissipation rate $\chi_{\Lambda Y_{R,eq}}$ (Eq. \ref{eq:chi-ly-sm}) and the terms in $\dot{m}_{Y_{R,eq}}$ (Eq. \ref{eq:source-term-y-ref-eq-sm}).

\begin{figure}[!htbp]
  \centering
  \includegraphics[width=0.6\textwidth]{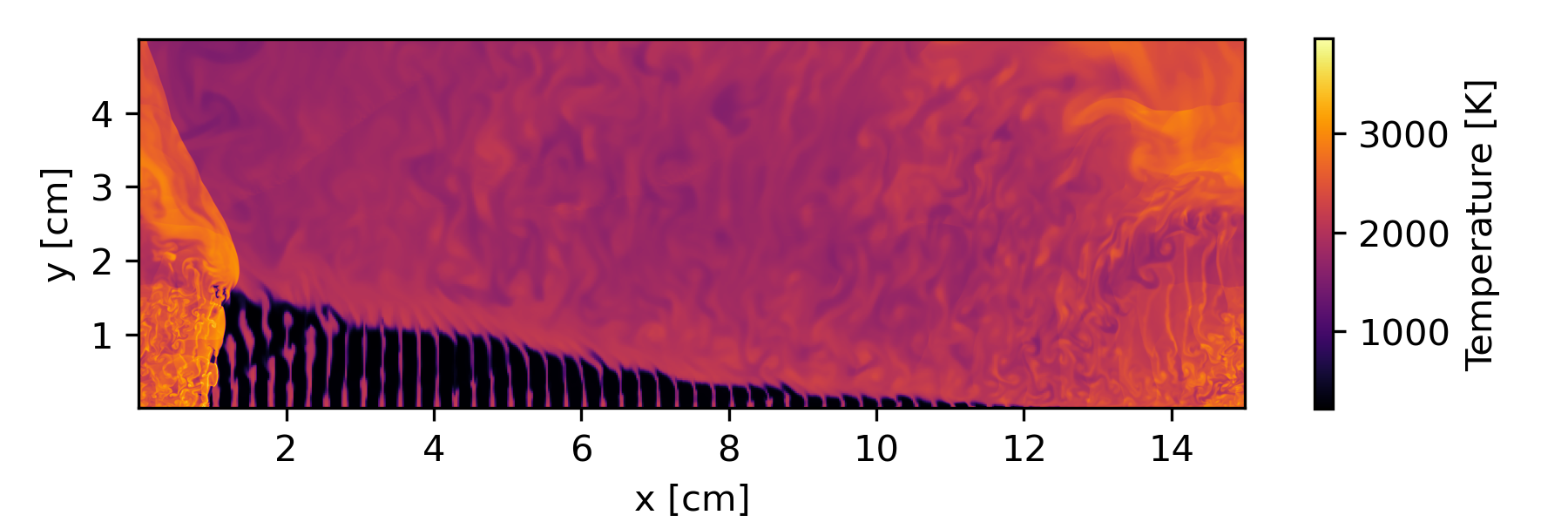}
  \caption{Temperature field in the RDE-like simulation at a snapshot after five wave cycles.}
  \label{Fig:rde-temperature}
\end{figure}

\begin{figure}[!htbp]
    \centering
    \includegraphics[width=0.8\linewidth]{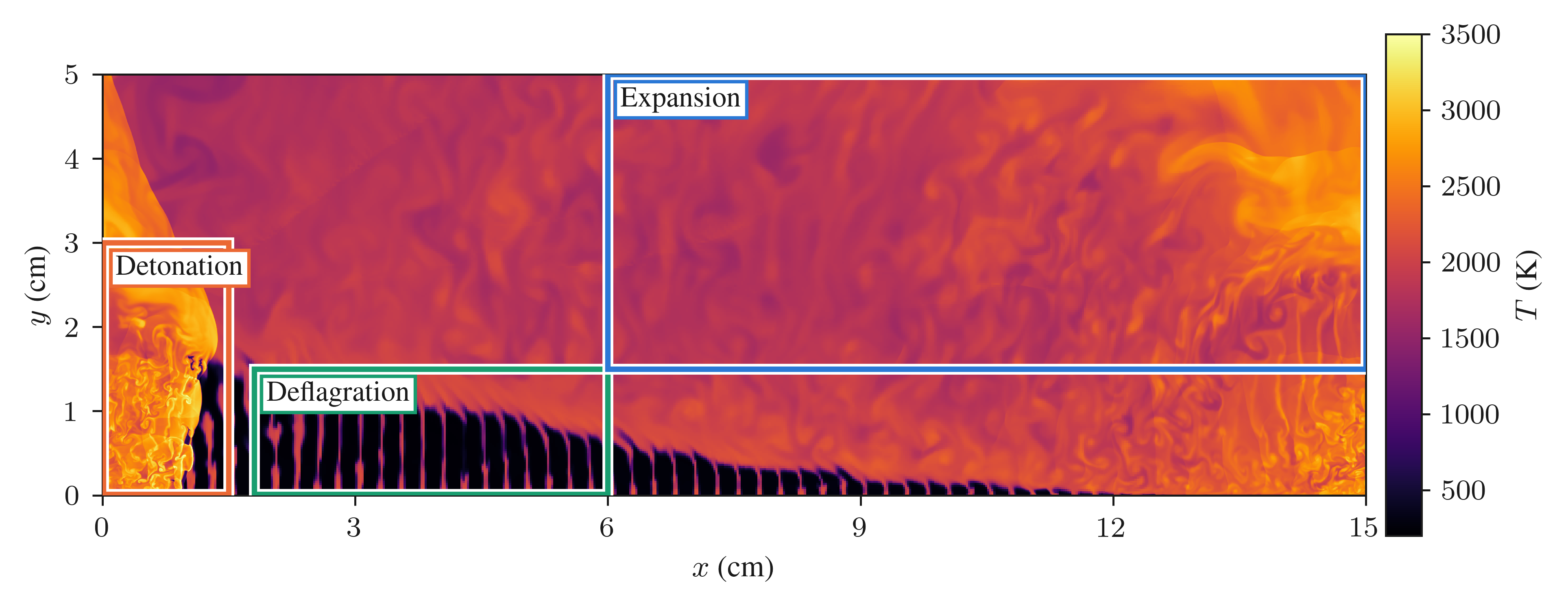}
    \caption{Temperature field of the RDE-like simulation shortly after the fifth wave cycle, with the
    detonation, deflagration (fill), and expansion regions used in Fig.~\ref{fig:term-compare} outlined.}
    \label{fig:regions}
\end{figure}

\begin{figure}[!htbp]
    \centering
    \includegraphics[width=0.7\linewidth]{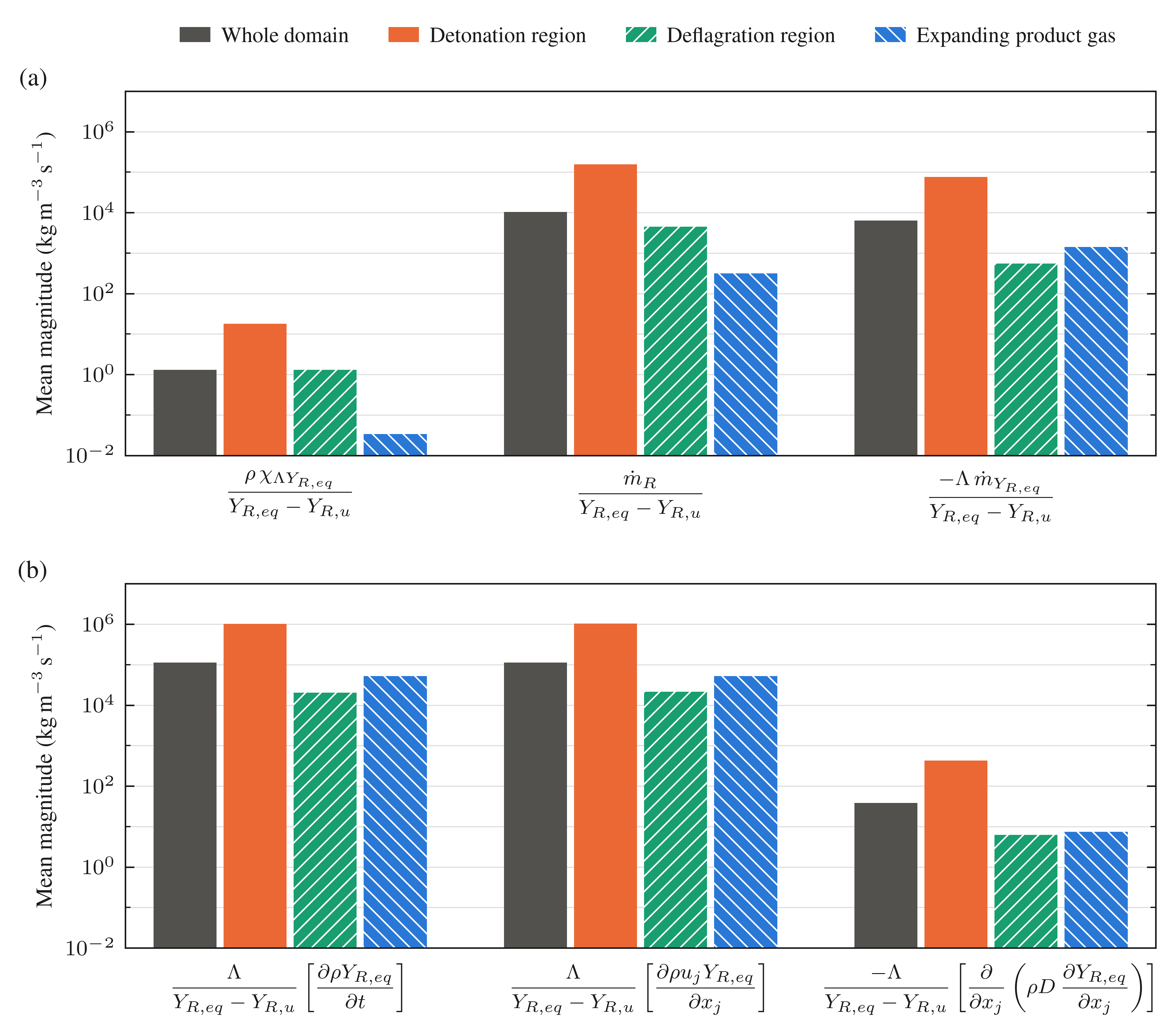}
    \caption{Mean magnitude of (a) the terms of $\dot{m}_{\Lambda,c}$ (Eq.~\ref{eq:high-speed-source-term-sm}) and
    (b) the $\Lambda$-weighted terms of $\dot{m}_{Y_{R,eq}}$ (Eq.~\ref{eq:source-term-y-ref-eq-sm}), over the whole
    domain and the three regions of Fig.~\ref{fig:regions}.}
    \label{fig:term-compare}
\end{figure}

To quantitatively assess their importance, the terms comprising Eq. \ref{eq:high-speed-source-term-sm} and Eq. \ref{eq:source-term-y-ref-eq-sm} were computed and spatially averaged over the domain at a given instant in time. The temperature field over the domain is displayed in Figs. \ref{Fig:rde-temperature} and \ref{fig:regions}, which displays the sub-regions selected for further analysis. Figure \ref{fig:term-compare} shows that $\rho \chi_{\Lambda Y_{R,eq}} / (Y_{R,eq} - Y_{R,u})$ is roughly four orders of magnitude smaller than the next smallest term. Focusing now on the terms in Eq. \ref{eq:source-term-y-ref-eq-sm}, Fig. \ref{fig:term-compare} demonstrates that the diffusion term for $Y_{R,eq}$ is about three orders smaller than the temporal and convection terms, which are the largest in magnitude and comparable to each other.

To ensure this analysis is applicable locally as well as globally, the analysis was similarly performed on characteristic regions: the detonation region, the fill region with parasitic deflagration, and the expanding product gas. The temperature fields of these particular subdomains are presented in Fig. \ref{fig:regions}. Although certain trends differed from the global analysis (e.g., $\dot{m}_{R}/(Y_{R,eq}-Y_{R,u})$ is smaller than $\Lambda \dot{m}_{Y_{R,eq}}/(Y_{R,eq}-Y_{R,u})$ in the expanding gas region), the overall trends regarding the terms of interest remain consistent across all regions. Specifically, the term $\rho \chi_{\Lambda, Y_{R,eq}} /(Y_{R,eq}-Y_{R,u})$ is always much smaller in magnitude than the other terms in Eq. \ref{eq:high-speed-source-term-sm} as seen in Fig. \ref{fig:term-compare}.

Similarly, the diffusion term in $\dot{m}_{Y_{R,eq}}$ is always several orders of magnitude smaller than the temporal term. Although this warrants further investigation in more diverse contexts before general conclusions can be made, this analysis suggests that these terms ($\chi_{\Lambda Y_{R,eq}}$ and $
\frac{\partial}{\partial x_j}
\left(
\rho D
\frac{\partial Y_{R,\mathrm{eq}}}{\partial x_j}
\right)$) can be assumed negligible, yielding Eq. \ref{eq:final-compressible-prog} in the main manuscript, which, following the modeling of the Lagrangian derivatives for $e$ and $\rho$, can then be computed from the model, convolved against the PDF, and used in the subsequent LES timestep to evolve the filtered compressible progress variable $\widetilde{\Lambda}$ and its variance $\Lambda_{v}$.

\FloatBarrier
\section{Subfilter Variance Analysis for the High-Fidelity Dataset}
\label{sec:variance-analysis}
Section \ref{flame-structure} in the manuscript analyzed the effect of the progress variable variance on the filtered thermochemical state for the one-dimensional detonation. An increase in the progress variable variance resulted in only a slight increase in the error.

To further quantify the influence of the progress variable variance, a similar analysis was performed on the data from the high-fidelty RDE simulation presented in Sec. \ref{rde-analysis}. In particular, the converged profiles at the points of maximum $\dot{m}_{\mathrm H_2 O}$ were selected from the extracted data in Figs. \ref{fig:rde-deflag-slice} and \ref{fig:rde-det-slice}. Following this, a clipped Gaussian filter ($G \propto \exp(-6r^2/\Delta^2)$ for $|r| \le \Delta$) was applied to the data to produce filtered inputs to the model representative of a filtered LES state. Then, the iterative algorithm was applied and convolved against the beta PDF to produce the model's filtered thermochemical state, which was then compared to the exact thermochemical state. This was performed with increasing width (and subsequently increasing variance) to analyze the role of the variance.

\begin{figure}[!htbp]
  \centering
  \includegraphics[width=0.7\textwidth]{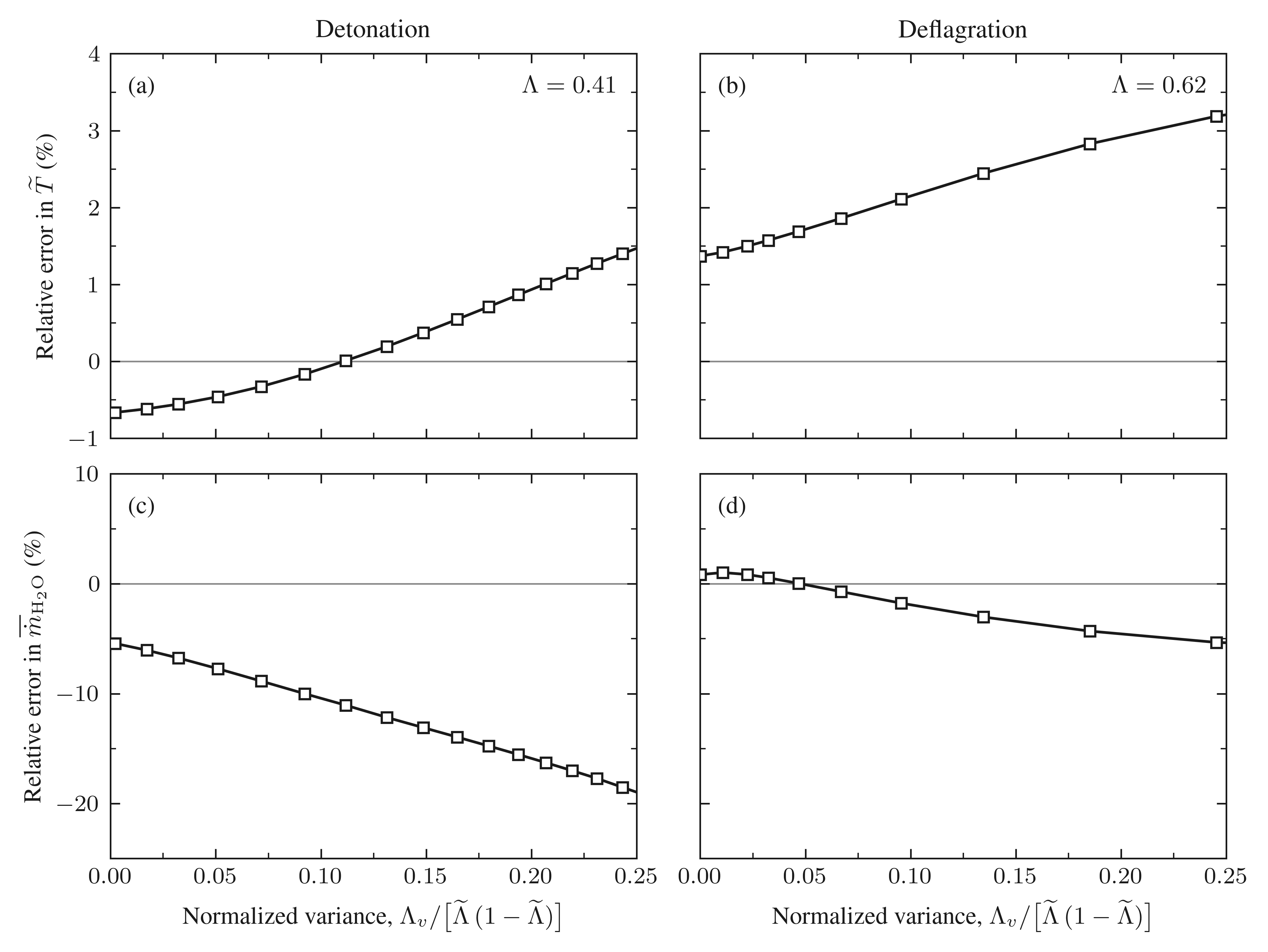}
  \caption{Relative error between the filtered thermochemical state and that obtained from the model using a beta PDF parameterized by the local $\widetilde{\Lambda}$ and $\Lambda_v$. (a,~c) Detonation: the RDE slice of Figs.~\ref{fig:rde-det-slice} and~\ref{fig:det-model-compare}. (b,~d) Deflagration: the RDE slice of Figs.~\ref{fig:rde-deflag-slice} and~\ref{fig:deflag-model-compare}. Relative error of (a,~b) the Favre-filtered temperature $\widetilde{T}$ and (c,~d) the filtered H$_2$O source term $\overline{\dot{m}}_{\mathrm{H_2O}}$.}
  \label{fig:rde-variance}
\end{figure}

In this analysis, when compared to that of Sec. \ref{flame-structure}, the variance played a larger role as seen in Fig. \ref{fig:rde-variance}. Figures \ref{fig:rde-variance}a and b display the relative error in $\widetilde{T}$, and Figs. \ref{fig:rde-variance}c and d display the error in $\overline{\dot{m}}_{\mathrm{H_2O}}$, while Figs. \ref{fig:rde-variance}a and c are for the detonation from Fig. \ref{fig:rde-det-slice} and Figs. \ref{fig:rde-variance}b and d are from the deflagration from Fig. \ref{fig:rde-deflag-slice}. As expected for the deflagrations, the relative error for both $\widetilde{T}$ and $\overline{\dot{m}}_{\mathrm{H_2O}}$ is mostly unaffected by the variance with errors staying within $10\%$ for almost all variances tested. The detonation accrues error with increasing variance which is reasonably small (under $5\%$) for $\widetilde{T}$; however, the source term errors approach $20\%$ as the variance reaches a large normalized variance of 0.25. It is worth pointing out that this error of $20\%$ is still considerably less than the $\approx80\%$ errors from the other models without considering filtering. This analysis suggests that, while the thermodynamically consistent model enables the application of LES to high-speed and supersonic compressible reacting flows, it is still prudent to apply AMR to reduce the variance in the cells containing the detonation.

\end{document}